\documentclass{emulateapj}
\usepackage{apjfonts}

\newcommand{\etal}{et al.}
\newcommand{\mbh}{M_{\rm BH}}
\newcommand{\mgal}{M_{\rm gal}}
\newcommand{\figzoom}{\epsscale{1.15}}
\newcommand{\plotter}{\plotone}

\shorttitle{Quasar and Elliptical Clustering}
\shortauthors{Hopkins \etal}
\slugcomment{Submitted to ApJ, November 26, 2006}
\begin{document}

\title{The Co-Formation of Spheroids and Quasars Traced in their Clustering}
\author{Philip F. Hopkins\altaffilmark{1}, 
Adam Lidz\altaffilmark{1},
Lars Hernquist\altaffilmark{1}, 
Alison L. Coil\altaffilmark{2}, 
Adam D. Myers\altaffilmark{3,4},
Thomas J. Cox\altaffilmark{1}, \&\
David N. Spergel\altaffilmark{5}
}
\altaffiltext{1}{Harvard-Smithsonian Center for Astrophysics, 60 Garden Street, Cambridge, MA 02138}
\altaffiltext{2}{Steward Observatory, University of Arizona, Tucson, AZ 85721}
\altaffiltext{3}{Department of Astronomy, University of Illinois at Urbana-Champaign, Urbana, IL 61801}
\altaffiltext{4}{National Center for Supercomputing Applications, Champaign, IL 61820}
\altaffiltext{5}{Dept. of Astrophysical Sciences, Peyton Hall, Princeton University, Princeton, NJ 08544}

\begin{abstract}

We compare observed clustering of quasars and galaxies as a 
function of redshift, mass, luminosity, and color/morphology, to constrain 
models of quasar fueling and the co-evolution of spheroids and supermassive
black holes (BHs).
High redshift quasars are shown to be drawn from the progenitors 
of local early-type galaxies, with the characteristic quasar 
luminosity $L_{\ast}$ at a given redshift reflecting a characteristic 
mass of the ``active'' BH/host population at that epoch. 
In detail, the empirically calculated clustering of quasar ``remnants'' (knowing the 
observed clustering of the original quasars) as a function of 
stellar mass and/or luminosity is identical to that observed 
for early-type populations. 
However, at a given redshift, the active quasars 
cluster as an ``intermediate'' population, reflecting neither ``typical'' 
late nor early-type galaxies at that redshift. 
Comparing with the age of elliptical 
stellar populations as a function of mass reveals that this 
``intermediate'' population represents those ellipticals 
undergoing (or terminating) their final significant star formation activity 
at the given epoch. Assuming that 
quasar triggering is associated with the formation/termination 
epoch of ellipticals predicts quasar 
clustering at all observed redshifts without any model dependence or 
assumptions about quasar light curves, lifetimes, or accretion rates. 
This is not true for spiral/disk populations or the quasar halos 
(by any definition of their ages); i.e.\ quasars do {\em not} generically 
trace star formation or halo 
assembly/growth processes. Interestingly, however, quasar clustering 
at all redshifts is consistent with a constant host halo mass 
$\sim4\times10^{12}\,h^{-1}\,M_{\sun}$, similar to the local ``group scale.''
The observations support scenarios in which 
major mergers dominate the bright, high-redshift quasar population. 
We demonstrate that future observation of quasar clustering 
as a function of luminosity can be used to constrain 
different fueling mechanisms which may dominate 
AGN populations at lower luminosity and/or redshift. 
We also show that clustering measurements at $z=3-6$ will be 
sensitive to the efficiency of feedback or ``quenching'' at these redshifts. 

\end{abstract}


\section{Introduction}
\label{sec:intro}

In recent years, it has become clear that essentially all 
galaxies harbor supermassive black holes (BHs) \citep[e.g.,][]{KR95}, 
the masses of which are correlated with many properties 
of their host spheroids, including luminosity, mass \citep{Magorrian98}, 
velocity dispersion \citep{FM00,Gebhardt00}, concentration 
and Sersic index \citep{Graham01}. 
Further, comparison of the relic BH mass density and quasar 
luminosity functions \citep[QLFs;][]{Soltan82,Salucci99,YT02,Marconi04,Shankar04,
HRH06}, 
the cosmic X-ray background \citep{ERZ02,Ueda03,Cao05,HRH06}, 
and relic BH Eddington ratios \citep{HHN06} demonstrates that 
the growth of BHs is dominated by a short-lived phase of high 
accretion rate, bright quasar activity. This, together with  
the similarity between the cosmic star formation history 
\citep[e.g.,][]{Hopkins06.sfh} and quasar accretion history 
\citep[e.g.,][]{Merloni04.hosts,HRH06}, 
reveals that quasar activity (with associated BH growth) and galaxy formation 
are linked. 

A number of theoretical models have been proposed to explain the 
evolution of these populations with redshift, and their 
correlations with one another 
\citep[][]{KH00,Somerville01,WyitheLoeb03,Granato04,Scannapieco04,
Baugh05,Monaco05,Croton05,H06b,H06c,H06d,Bower06,deLucia06,Malbon06}. In many of these 
models, the merger hypothesis \citep{Toomre77} provides 
a physical mechanism linking galaxy star formation, morphology, and 
black hole evolution. According to
this picture, gas-rich galaxy mergers channel 
large amounts of gas to galaxy centers
\citep[e.g.,][]{BH91,BH96}, fueling powerful starbursts \citep[e.g.,][]{MH96}
and buried BH growth \citep[][]{Sanders88} until the BH
grows large enough that feedback from accretion rapidly unbinds and heats the
surrounding gas \citep{SR98}, briefly revealing a bright, optical quasar 
\citep{H05a}. As gas densities and corresponding accretion rates 
rapidly decline, an inactive ``dead'' BH is left in an elliptical galaxy
satisfying observed correlations between BH and spheroid mass. 
Major mergers rapidly and efficiently exhaust the cold gas reservoirs of the 
progenitor systems, allowing the remnant to rapidly redden with a low 
specific star formation rate, with the process potentially accelerated by the 
expulsion of gas by the quasar \citep[][]{SDH05a}. 

Recent hydrodynamical simulations, incorporating star formation, supernova
feedback, and BH growth and feedback \citep{SDH05b} make it possible
to study these processes dynamically and have lent support to this 
general scenario.  Mergers with BH feedback
yield remnants resembling observed ellipticals in their correlations
with BH properties \citep{DiMatteo05}, scaling relations
\citep{Robertson06b}, colors \citep{SDH05a}, and morphological and
kinematic
properties \citep{Cox06}.  The quasar activity excited through such
mergers can account for the QLF and a wide range of quasar properties
at a number of frequencies \citep{H05a,H06b}, and with such a detailed
model to ``map'' between merger, quasar, and remnant galaxy
populations it is possible to show that the buildup and statistics of
the quasar, red galaxy, and merger 
mass and luminosity functions are consistent and can be used
to predict one another \citep{H06c,H06d,H06e}.

However, it is by no means clear whether this is, in fact, the dominant 
mechanism for the triggering of quasars and buildup of early-type populations. 
For example, the association between BH and stellar mass 
discussed above leads some models to tie quasar activity directly 
to star formation \citep[e.g.,][]{Granato04}, implying it will evolve in a manner 
tracing star-forming galaxies, with this evolution and the corresponding downsizing 
effect roughly independent of mergers and morphological 
galaxy segregation at redshifts $z\lesssim2$. 
Others invoke post-starburst AGN feedback to suppress 
star formation on long timescales and at relatively low 
accretion rates through, e.g.,\ ``radio-mode'' feedback 
\citep{Croton05}. In this specific case, the ``radio-mode'' is associated with 
low-luminosity activity after a quasar phase builds a massive BH (i.e.\ quasar ``relics''), but 
it is possible to construct scenarios \citep[e.g.,][]{CiottiOstriker01,Binney04} in which 
the same task is accomplished by cyclic, potentially ``quasar-like'' (i.e.\ 
high Eddington ratio) bursts of activity, or in which the ``radio-mode'' 
might be directly associated with an optically luminous ``quasar mode,''
either of which would imply quasars should trace the established ``old'' red galaxy 
population at each redshift. 
In several models, a distinction between ``hot'' and ``cold'' accretion modes 
\citep{Birnboim03,Keres05,Dekel06}, in which new gas cannot cool into a 
galactic disk above a critical dark matter halo mass, determines the 
formation of red galaxy populations, essentially independent of 
quasar triggering \citep[e.g.,][but see also Binney 2004]{Cattaneo06}. 

At low luminosities, ($M_{B}\gtrsim-23$, important at $z\lesssim0.5$), 
models predict that stochastic, high-Eddington ratio ``Seyfert'' activity 
triggered in gas-rich, disk-dominated systems 
will contribute increasingly to the AGN luminosity 
function \citep{HH06}, with enhancements to these fueling 
mechanisms from bar instabilities and galaxy harassment. Indeed, 
the morphological makeup of low-luminosity, low-redshift 
Seyferts appears to support this, with increasing dominance of 
unperturbed disks at low luminosities seen locally 
\citep[e.g.,][]{Kauffmann03.agnhosts,DongDeRobertis06} and 
at low redshift $z\sim0-1$ \citep{Sanchez04,Pierce06}. 
At high luminosities 
(even at these redshifts), however, the quasar populations 
are increasingly dominated by ellipticals and merger 
remnants, particularly those with young 
stellar populations suggesting recent starburst 
activity \citep[][]{Kauffmann03.agnhosts,Sanchez04,
VandenBerk06,Best05,DongDeRobertis06}, 
and even clear merger remnants \citep{Sanchez04}. 
Still, some models extend the observed fueling in disk systems 
to high redshift quasars, invoking disk instabilities in 
very gas-rich high redshift disks as a primary
triggering mechanism \citep{KH00}. 

Clearly, observations which can break the degeneracies 
between these quasar fueling models are of great interest. 
Unfortunately, comparison of the quasar and galaxy or host luminosity 
functions, while important, suffers from a number of degeneracies 
and can be ``tuned'' in most semi-analytic models. 
Direct observations of host morphologies, while an ideal tool 
for this study, are difficult at high redshift and 
highly incomplete, especially for 
bright quasars which dominate their host galaxy light 
in all observed wavebands. However, the clustering of these 
populations may represent a robust test of their 
potential correlations, which does not depend sensitively 
on sample selection. Critically, considering the clustering 
of quasars and their potential hosts is not highly 
model-dependent in the way of, e.g.,\ mapping between 
their luminosity functions or modeling their triggering 
rates in an a priori fashion.

In recent years, wide-field surveys such as 
the Two Degree Field (2dF)
QSO Redshift Survey \citep[2QZ;][]{Boyle00} and the Sloan Digital Sky
Survey \citep[SDSS;][]{York00} have enabled tight 
measurements of quasar clustering to redshifts $z\sim3$, and 
a detailed breakdown of galaxy clustering as a function of 
galaxy mass, luminosity, color, and morphology 
\citep[e.g.,][]{Norberg02,Zehavi02,Li06}. These observations 
allow us to consider the possible triggering mechanisms 
of quasars in a robust, empirical manner, and answer 
several key questions. Which local populations have the 
appropriate clustering 
to be the descendants of high-redshift quasars? How is 
the quasar epoch of these populations related to
galaxy formation? And, to the extent that quasars are associated 
with spheroid formation, are bright quasar populations 
dominated by quasars triggered in formation ``events''?

In this paper, we investigate the link between 
quasar activity and galaxy formation by comparing the observed clustering 
of quasar and galaxy populations as a function of 
mass, luminosity, color, and redshift. In \S~\ref{sec:local}, we 
compare the clustering of quasars and local galaxies to 
determine which galaxy populations ``descend'' from 
high-redshift quasar progenitors. In \S~\ref{sec:bias.vs.lum} 
we consider the clustering of quasars as a function of luminosity 
and redshift, checking the robustness of our results and presenting 
tests for the dominance of different AGN fueling mechanisms 
at low luminosities. \S~\ref{sec:populations} compares the clustering 
of quasars as a function of redshift with that of different galaxy populations 
at the same redshift, 
ruling out several classes of fueling models. \S~\ref{sec:ages} 
further considers the age as a function of stellar and BH mass of these 
galaxy populations, and uses this to predict quasar clustering 
as a function of redshift for different host populations. 
In \S~\ref{sec:redshift}, we use these 
comparisons to predict quasar clustering at high redshifts, 
presenting observational tests to determine the efficiency of 
high-redshift quasar feedback. Finally, in \S~\ref{sec:discussion} 
we discuss our results and conclusions, and their implications 
for various models of quasar triggering and BH-spheroid 
co-evolution. 

Throughout, we adopt a WMAP1 
$(\Omega_{\rm M},\,\Omega_{\Lambda},\,h,\,\sigma_{8},\,n_{s})
=(0.27,\,0.73,\,0.71,\,0.84,\,0.96)$ cosmology 
\citep{Spergel03}, and normalize all observations and models 
shown to this cosmology. Although the exact choice of 
cosmology may systematically 
shift the inferred bias and halo masses (primarily scaling with $\sigma_{8}$), 
our comparisons (i.e.\ relative biases) are unchanged, 
and repeating our calculations for 
a ``concordance'' $(0.3,\,0.7,\,0.7,\,0.9,\,1.0)$ cosmology or 
the WMAP3 results of Spergel et al.\ (2006)
has little effect on our conclusions. 
All magnitudes are in the Vega system.

\section{Using Clustering to Determine the Parent Population of Quasars/Ellipticals}
\label{sec:local}

At a given redshift $z_{i}$, quasars are being triggered in some ``parent'' halo population. 
These halos, and by consequence the quasars they host, cluster with 
some bias/amplitude $b(z_{i})$. 
The halos will subsequently evolve via gravitational clustering, which in 
linear theory predicts their subsequent clustering at any later 
redshift $z_{f}$ will be given by 
\begin{equation}
b(z_{f}) = 1 + \frac{D(z_{i})}{D(z_{f})}\,{\Bigl[}b(z_{i})-1{\Bigr]}
\label{eqn:bias.evol}
\end{equation}
\citep{Fry96,MoWhite96,Croom01}, 
where $D(z)$ is the linear growth factor. 
Thus, at $z=0$, the halos which hosted the quasars at $z_{i}$ will 
have a bias of $b(0)=1+D(z_{i})\,[b(z_{i})-1]$. 

The quasar luminosity function at a given redshift has a 
characteristic luminosity $L_{\ast}$. 
Given that quasars (at least those with $L\gtrsim L_{\ast}$) 
are typically observed to have high Eddington 
ratios $\lambda\equiv L/L_{\rm Edd}\approx 0.3-0.5$
\citep[][]{Heckman04,Vestergaard04,MD04,Kollmeier05}, this 
$L_{\ast}$ reflects the characteristic mass 
of ``active'' BHs at that redshift, 
$M_{\rm BH}\approx3.0\,\lambda^{-1}\times10^{8}\,M_{\sun}\,(L_{\ast}/10^{13}\,L_{\sun})$.
Direct observations of quasar Eddington ratios/BH masses \citep{MD04,Kollmeier05}, 
limits from the X-ray background \citep[][]{ERZ02,Cao05} and 
BH mass functions \citep[][]{Soltan82,YT02,Marconi04,Shankar04,HRH06},
radio luminosity functions 
\citep{Merloni04.bhmf,McLure04}, and ``relic'' low-luminosity Eddington 
ratios \citep{HHN06} all 
rule out the possibility that these BHs subsequently gain 
significant mass ($\gtrsim10-20\%$ growth) after their brief ``active'' phase, 
so the $M_{\rm BH}$ above is equivalent (within its errors) to the 
$z=0$ BH mass of these objects. 
Since the relationship between BH mass and host luminosity or 
mass is well-determined at $z=0$
($M_{\rm BH}=\mu\,\mgal$ with $\mu\approx0.001$; Marconi \& Hunt 2003, H{\"a}ring \& Rix 2004), knowing $M_{\rm BH}(z=0)$ of a population implies, with 
little uncertainty, its $z=0$ host mass $\mgal$ or luminosity. 

\begin{figure*}
    \centering
    \figzoom
    \plotone{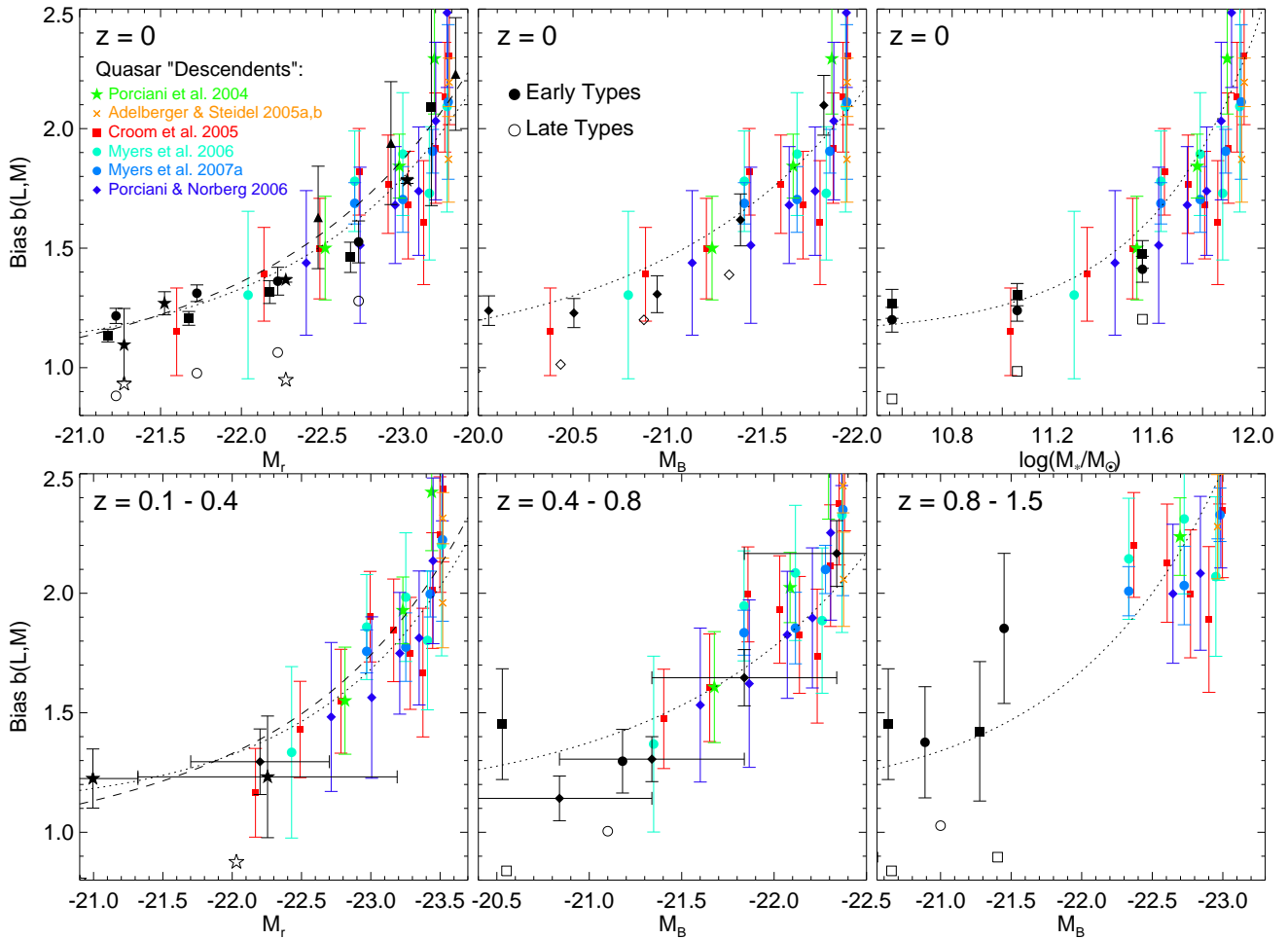}
    \caption{Evolved clustering of quasar ``descendents'' (colored points) as a function of 
    mass or luminosity, compared to clustering of 
    early type (solid black points) and late type (open black points) 
    galaxies of the same mass/luminosity. The measured clustering of quasars at a given $z$ 
    (samples as labeled)
    is evolved in linear theory to the given observed redshift, and plotted 
    as of function of the relic host galaxy mass/luminosity. Galaxy clustering 
    is shown at $z=0$ from 
    \citet[][color and morphologically-selected early types as squares and circles, respectively]{Li06}, 
    \citet[][color-selected; stars]{Zehavi05}, \citet[][SDSS LRGs; triangles]{Percival06}, 
    \citet[][diamonds]{Norberg02}; 
    at $z=0.1-0.4$ from 
    \citet[][stars]{Shepherd01} and \citet[][and references therein; diamond]{Brown03};  
    at $z=0.4-0.8$ from 
    \citet[][square]{Meneux06}, \citet[][circle]{Phleps05}, 
    and \citet[][and references therein; diamonds]{Brown03};  
    and at $z=0.8-1.5$ from 
    \citet[][square]{Meneux06}, and
    \citet[][circles]{Coil04.z1clustering}. 
    Fitted lines show the best-fit bias of early type galaxies at each $z$ 
    as a function of mass/luminosity 
    \citep{Norberg01,Tegmark04,Li06}.
    \label{fig:local.cluster.compare}}
\end{figure*}

In Figure~\ref{fig:local.cluster.compare}, we consider the clustering of quasars 
as a function of redshift, evolved to $z=0$. At each redshift where the quasar clustering 
$b(z)$ is measured, we also know the characteristic luminosity 
$L_{\ast}$. Here we adopt the bolometric $L_{\ast}$ determined in the 
observational compilation of 
\citet{HRH06}, 
\begin{eqnarray} 
 & \log(L_{\ast}/L_{\sun}) = k_{0} + k_{1}\,\xi + k_{2}\,\xi^{2} + k_{3}\,\xi^{3}, & \\
 & \xi\equiv\log{{\Bigl(}\frac{1+z}{1+z_{\rm ref}}{\Bigr)}}\ \ \ \ \ \ [z_{\rm ref}=2], & 
\label{eqn:lstar}
\end{eqnarray} 
($k_{0}=13.036$; $k_{1}=0.632$; $k_{2}=-11.76$; $k_{3}=-14.25$), 
but for our purposes this is identical to adopting the 
$B$-band or X-ray $L_{\ast}$ from \citet{Ueda03,Croom04,HMS05,Richards06} 
and converting it to a bolometric $L_{\ast}$ with a typical 
bolometric correction from \citet{Marconi04,Richards06.seds,HRH06}. 
Given the conversions above, we consider the implied characteristic BH mass 
and, assuming little subsequent BH growth, the corresponding 
$z=0$ stellar mass or luminosity in a given band 
\citep[here from the $M_{\rm BH}-L_{\rm host}$ relations of][]{MarconiHunt03}. 
Knowing how the bias of these halos evolves to $z=0$ (Equation~\ref{eqn:bias.evol}), 
we plot the 
bias as a function of stellar mass, at $z=0$, of the evolved quasar ``parent'' population. 
We compare this with observed bias as a function of stellar mass or luminosity 
for both early and late-type galaxies.  Note that
the lowest and highest-redshift 
bins ($z\sim0.5$ and $z\sim2.5$, respectively) 
in the \citet{Myers06b} quasar clustering measurements are 
significantly affected by catastrophic redshift errors (owing to their considering all 
photometrically classified quasars, as opposed to just spectroscopically confirmed quasars); 
we follow their suggestion and decrease (increase) the clustering 
amplitude in the lowest (highest) redshift bin by $\sim20\%$; excluding these points 
entirely, however, has no effect on our conclusions. 

\begin{figure}
    \centering
    \figzoom
    \plotone{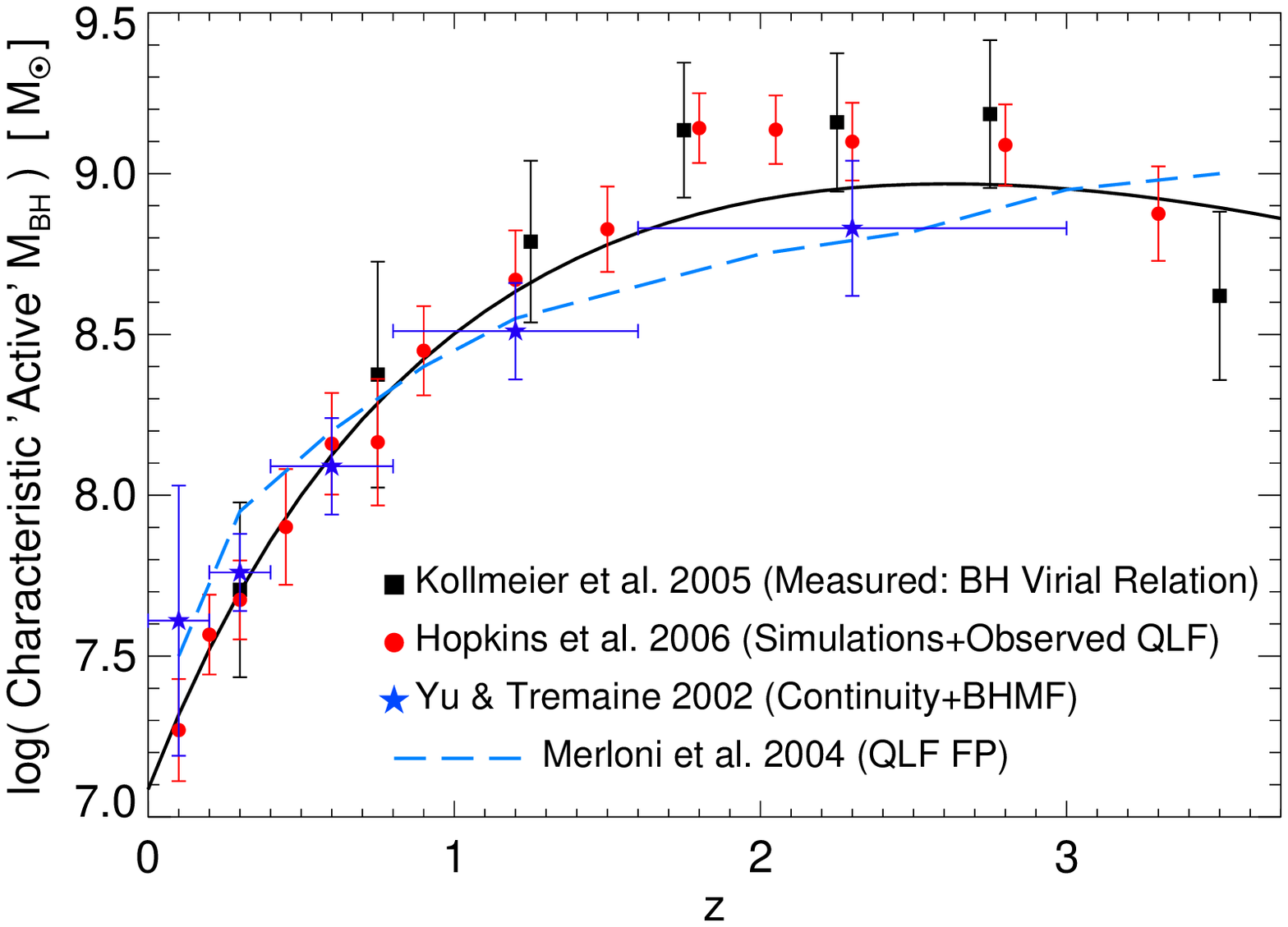}
    \caption{Characteristic ``active'' mass of BHs at a given redshift, i.e.\ 
    the BH mass corresponding to (dominant at) $L_{\ast}$ in the QLF. 
    Black squares adopt the virial relation BH mass 
    determinations of \citet{Kollmeier05}. Red circles fit the observed 
    QLF at each redshift \citep{HRH06} to quasar light curve models \citep{H06b}. 
    Blue stars adopt the simplified continuity model 
    from \citet{YT02} and \citet{Marconi04} given the \citet{Ueda03} QLF. Blue dashed line 
    \citep{Merloni04.bhmf} and derives the active mass distribution from
    the radio and X-ray ``black hole fundamental plane.'' Black solid line is a 
    fitted relation (Equations~(\ref{eqn:lstar}) \&\ (\ref{eqn:bestfit.bias})). 
    The ``active'' BH mass is 
    well-defined. Adopting different estimators does not 
    significantly alter our conclusions or comparisons in Figures~\ref{fig:local.cluster.compare} 
    \&\ \ref{fig:age.bias.compare}. 
    \label{fig:active.mbh}}
\end{figure}

For reference, we show the characteristic $M_{\rm BH}$ corresponding 
to $L_{\ast}$ in the QLF (as 
defined above) as a function of redshift in Figure~\ref{fig:active.mbh}. 
Whether we adopt a direct conversion from the observed $L_{\ast}$ 
(Equation~(\ref{eqn:lstar})) with observed Eddington ratios, as above, 
or invoke any of several empirical models for the QLF Eddington ratio 
distribution, we obtain a similar $M_{\rm BH}$. The figure illustrates the 
inherent factor $\lesssim2$ 
systematic uncertainty in the appropriate Eddington ratios 
and bolometric corrections used in our conversions. These uncertainties, 
however, are at most comparable to the uncertainties in 
quasar clustering measurements. Because of this, our conclusions and 
comparisons are not sensitive to the exact method we use to 
estimate the $M_{\rm BH}$ corresponding to $L_{\ast}$ at a given redshift.

The comparison in Figure~\ref{fig:local.cluster.compare} 
is possible at any redshift, not simply $z=0$. 
We repeat our methodology above at several $z_{\rm obs}=0-1$, 
evolving the bias to $b(z_{\rm obs})$. The agreement with 
red galaxy clustering observed as a function of mass at each $z_{\rm obs}$ 
is good, at all $z\lesssim1$. At higher redshifts, small fields in galaxy 
surveys limit one's ability to measure clustering as a bivariate function of 
luminosity and color/morphology at the highest luminosities, where the relics of 
$z\sim2-3$ quasars are expected.

\section{Clustering as  Function of Luminosity and Different 
AGN Fueling Mechanisms}
\label{sec:bias.vs.lum}

The comparison in Figure~\ref{fig:local.cluster.compare} has one 
important caveat. We assumed that measurements of quasar 
clustering at a given redshift are representative of a ``characteristic'' 
active mass $M_{\rm BH}\propto L_{\ast}$ of the QLF. In other words, 
quasar clustering should be a weak function of the exact quasar 
luminosity, at least near $L_{\ast}$. If this 
were not true, our comparison would break down on two levels. First, 
it would be sensitive to the exact luminosity distribution of observed 
quasars. 
Second, if quasars of slightly different luminosities at the 
same redshift 
represented different BH/host masses (consequently 
making quasar clustering a strong function of quasar luminosity), there would 
be no well-defined ``characteristic'' active mass at that redshift. 

Fortunately, \citet{Lidz06} considered this question in detail, and 
demonstrated that realistic quasar light curve 
and lifetime models like those of
\citet{H06b} indeed predict a relatively flat 
quasar bias as a function of luminosity, in contrast to 
more naive models which assume a one-to-one correlation 
between observed quasar luminosity, BH mass, and host stellar/halo 
mass. This appears to be increasingly confirmed by direct observations, 
with \citet{Adelberger05.lifetimes,Croom05,Myers06,Myers06b,Porciani06,Coil06.agn} 
finding no evidence for a significant dependence of quasar clustering 
on luminosity.

\begin{figure*}
    \centering
    \figzoom
    \plotone{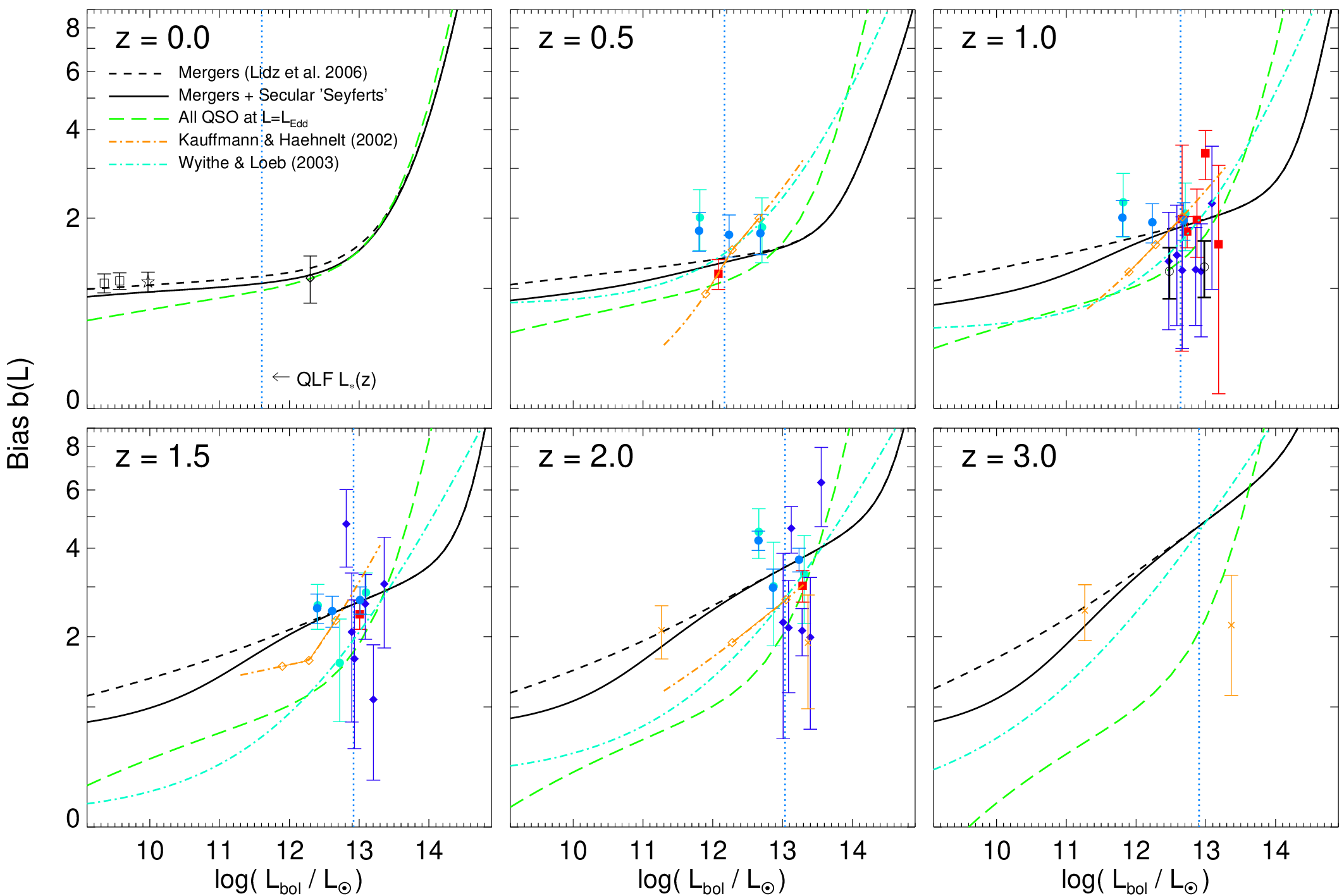}
    \caption{Quasar bias as a function of quasar luminosity 
    at each of several redshifts. Points show quasar observations from 
    Figure~\ref{fig:local.cluster.compare} (same style). Open black points 
    add the local ($z\lesssim0.3$) observations of 
    \citet[][SDSS LLAGN (LINERS+Seyferts); squares]{Constantin06}, 
    \citet[][SDSS; star]{Wake04}, and \citet[][AERQS; diamond]{Grazian04}, 
    and the $z\sim1$ cross-correlation measurements of 
    \citet[][SDSS+DEEP2; circles]{Coil06.agn}. Blue dotted line shows 
    $L_{\ast}(z)$. These are compared to various models described in 
    the text (curves, as labeled). Models in black adopt the feedback-regulated 
    quasar light curve/lifetime models from \citet{H06a}, others consider 
    more simplified ``light bulb'' model light curves. 
    \label{fig:bias.vs.lum}}
\end{figure*}

Figure~\ref{fig:bias.vs.lum} explicitly considers the dependence of 
bias on luminosity and its possible effects on our conclusions. 
We plot, at each of several redshifts, the observed bias of quasars 
as a function of luminosity. For the sake of direct comparison, 
all observations are converted to a bolometric luminosity 
with the bolometric corrections from \citet{HRH06}. The QLF 
break luminosity $L_{\ast}$ at each redshift, estimated in 
\citet{HRH06}, is also shown. The first thing to note is that 
the quasar observations with which we compare generally
sample the QLF very near $L_{\ast}$, so {\em regardless} of the 
dependence of bias on luminosity, our conclusions are 
not changed. We have, for example, recalculated the results of 
\S~\ref{sec:local} assuming that the characteristic mass 
of active BHs is given by $M_{\rm Edd}({\langle}L_{\rm obs}{\rangle})$, 
where ${\langle}L_{\rm obs}{\rangle}$ is the mean (or median) 
observed quasar luminosity in each clustering sample in 
Figure~\ref{fig:local.cluster.compare}, and find 
it makes no difference (changing the comparisons by $\ll 1\sigma$). 

We compare the observations with various theoretical models in Figure~\ref{fig:bias.vs.lum}. 
The models of \citet{H05b} define the conditional quasar lifetime;
i.e.\ time a 
quasar with a given final (relic) BH mass (or equivalently, peak quasar 
luminosity) spends over its lifetime in various luminosity intervals, 
$t_{Q}(L\,|\,M_{\rm BH})$. Since this is much less than the Hubble time at 
all redshifts of interest, the observed QLF ($\phi_{Q}[L]$) is given by the convolution 
of $t_{Q}(L\,|\,M_{\rm BH})$ with the rate at which quasars of a 
given relic mass $M_{\rm BH}$ are ``triggered'' or ``turned on,''
\begin{equation}
\phi_{Q}(L)=\int t_{Q}(L\,|\,M_{\rm BH})\,\dot{\phi}(M_{\rm BH})\,{\rm d}\log{M_{\rm BH}}, 
\label{eqn:model}
\end{equation}
where $\dot{\phi}(M_{\rm BH})$ is the rate of triggering, i.e.\ number of 
quasars formed or triggered per unit time per unit volume per 
logarithmic interval in relic mass. The integrand here defines the relative 
contribution to a given observed luminosity interval from 
each interval in $M_{\rm BH}$. Given the 
BH-host mass relation, we can convert this 
to the relative contribution from hosts of different masses 
$\mgal$ (i.e.\ ${\rm d}\phi_{Q}(L)/{\rm d}\log{\mgal}$). 
In detail, we assume $P(\mbh\,|\,\mgal)$ is 
distributed 
as a lognormal about the mean correlation, with a dispersion of 
$0.3$\,dex taken from observations \citep{MarconiHunt03,HaringRix04,Novak06} 
and hydrodynamical simulations \citep{DiMatteo05,Robertson06a}. 
Calculating the bias for a given relic $\mbh$ or $\mgal$ and observed redshift
as in \S~\ref{sec:local}, we can integrate over these contributions to 
determine the appropriately weighted mean bias as a function of 
observed quasar luminosity, 
\begin{equation}
{\langle}b{\rangle}(L,\,z)=\frac{1}{\phi_{Q}(L)}\,\int b(\mgal,\,z)\,\frac{{\rm d}\phi_{Q}(L)}{{\rm d}\log{\mgal}}\,{\rm d}\log{\mgal}.
\end{equation}

Although binning 
by both luminosity and redshift greatly reduces the size of observed samples 
and increases their errors, the observations in Figure~\ref{fig:bias.vs.lum} 
confirm the predictions of \citet{Lidz06} to the extent that they currently probe. 
To contrast, we construct an alternative ``straw-man'' model. 
Specifically, we compare 
with the naive expectation, if all quasars were at the same Eddington ratio 
(so-called ``light-bulb'' models),
i.e.\ if there was a one-to-one correlation between observed luminosity 
and BH mass (and correspondingly, host mass), which produces a 
much steeper trend of bias as a function of luminosity and 
is significantly disfavored ($>4.5\,\sigma$; although from any 
individual sample the significance is only $\sim2\,\sigma$). We also 
compare the 
predicted clustering as a function of luminosity from the semi-analytic 
models of \citet{KH02} and 
\citet{WyitheLoeb03}, which adopt idealized, strongly peaked/decaying exponential quasar 
light curves (i.e.\ Eddington-limited growth to a peak luminosity, then 
subsequent $L\propto \exp{(-t/t_{Q})}$) and therefore yield similar predictions to the 
constant Eddington ratio ``light bulb'' model (and are likewise disfavored at $>4\,\sigma$). 

\begin{figure}
    \centering
    \figzoom
    \plotone{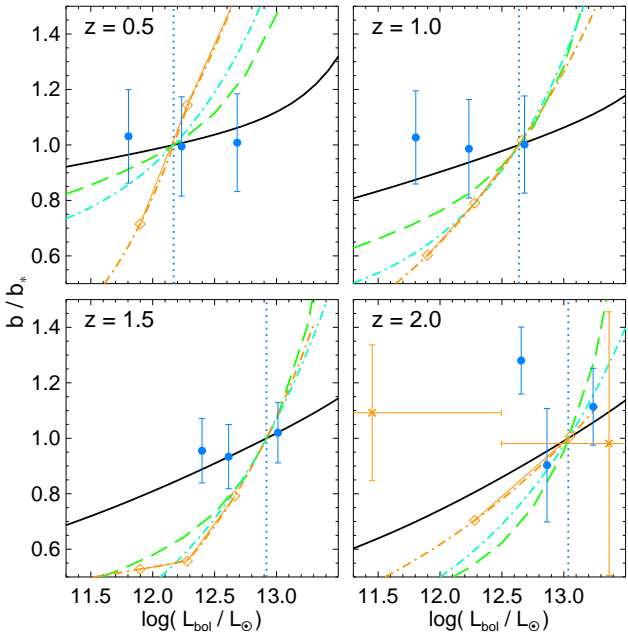}
    \caption{As Figure~\ref{fig:bias.vs.lum}, but showing the relative quasar bias 
    $b/b_{\ast}$ ($b_{\ast}\equiv b(L_{\ast})$) as a function of luminosity near $L_{\ast}$ 
    for the models and most well-constrained 
    \citep{Myers06b} and largest luminosity baseline \citep{Adelberger05.lifetimes} observations, 
    to highlight the luminosity dependence and differences between the models.
    \label{fig:bias.vs.lum.zoom}}
\end{figure}

\begin{figure}
    \centering
    \figzoom
    \plotone{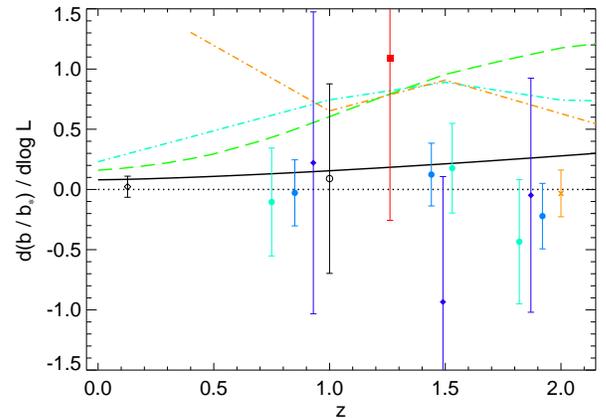}
    \caption{The best-fit dependence of quasar bias on luminosity (${\rm d}(b/b_{\ast})/{\rm d}\log{L}$) 
    from the observations in Figures~\ref{fig:bias.vs.lum} \& \ref{fig:bias.vs.lum.zoom} (points), 
    compared to the dependence expected from the models (calculated at $L_{\ast}$). 
    The observations favor little or no dependence of clustering on luminosity.   
    \label{fig:bias.lum.slope}}
\end{figure}

Figure~\ref{fig:bias.vs.lum.zoom} highlights the dependence of bias on luminosity
in the observations and models by plotting the relative bias $b/b_{\ast}$ 
(where $b_{\ast}\equiv b(L_{\ast})$) near the QLF $L_{\ast}$, more clearly 
demonstrating the observational indication of a weak dependence. Alternatively, we 
can fit each observed sample binned by luminosity at a given redshift to a 
``slope,''
\begin{equation}
\frac{b}{b_{\ast}} = 1 + \frac{{\rm d}(b/b_{\ast})}{{\rm d}\log{L}}\,\log{(L/L_{\ast})}, 
\end{equation}
the results of which are shown in Figure~\ref{fig:bias.lum.slope} as a function of 
redshift, compared to the slope (evaluated at $L_{\ast}$) predicted by the 
various models. At all redshifts, the observations are consistent with no 
dependence of clustering on luminosity, and strongly disfavor the 
``light-bulb'' class of models (again, at $\sim4\,\sigma$ at $z\sim1.5-2$). 
This confirms the conclusions of these studies individually, particularly the most recent observations 
from \citet{Myers06b} and the largest luminosity baseline observations from \citet{Adelberger05.lifetimes}.
The weak dependence predicted by the models of \citet{H06b,Lidz06} 
provides a considerably improved fit, although even it may be marginally too 
steep relative to the observations.

Galaxy clustering (and therefore, presumably, host halo mass) appear to be 
much more strongly correlated with galaxy luminosity or stellar mass 
(Figure~\ref{fig:local.cluster.compare}) than with quasar luminosity 
(at a given redshift); i.e.\ the weak dependence of bias on quasar luminosity 
appears to be driven 
by variation in Eddington ratios at a characteristic ``active'' mass. 
This is also supported by comparison of quasar luminosity functions and number counts, in a 
semi-analytic context \citep{Volonteri06}.
We note that this is completely consistent with observations 
that find similar high Eddington ratios 
for all bright quasars, as these are confined to $L\gtrsim L_{\ast}$, 
(and indeed the \citet{H06b,Lidz06} model predictions do, at these 
highest luminosities, reproduce this and 
imply a steep trend of bias with luminosity). 
However, the relatively weak trend in clustering 
near and below $L_{\ast}$ makes our 
conclusions throughout considerably more robust, so long as the observed quasar 
sample resolves $L_{\ast}$ (true for all plotted points).

Despite the detail of the models involved, the predictions in Figure~\ref{fig:bias.vs.lum} 
are all simplified in that they model only one mechanism for quasar fueling. 
However, \citet{HH06} (among others) predict that at low luminosities, 
contributions from smaller BHs in non-merging disk bulges, triggered 
by disk and bar instabilities, stochastic accretion, 
harassment, or other perturbations, are expected to 
dominate the ``Seyfert'' population. We therefore repeat our calculation, 
but allow different fueling mechanisms in different hosts to contribute 
to different quasar luminosities, according to the 
models of \citet{HH06} and \citet{Lidz06}. Because the ``Seyferts'' 
in this particular model \citep{HH06} are generally less massive 
systems at high Eddington ratio in blue, star-forming galaxies, they are less 
biased than merger remnants of similar observed luminosity.

The inclusion of 
these populations in Figure~\ref{fig:bias.vs.lum} does not 
change our conclusions near $L\sim L_{\ast}$. However it does introduce a 
feature, generally a sharp decrease in observed bias, at the 
luminosity where these secular fueling mechanisms begin to dominate the AGN 
population. This luminosity is typically quite low, $L\sim10^{11}-10^{12}\,L_{\sun}$ 
(corresponding roughly to luminosities below the classical 
Seyfert-quasar division of $M_{B}=-23$). The only redshift at which the clustering of 
such very low-luminosity AGN has been measured is $z\lesssim0.2$, 
by which point massive, gas-rich mergers 
are sufficiently rare that the predicted ``Seyfert'' population 
from \citet{HH06} dominates the merger-triggered quasar population 
at all luminosities, erasing the feature indicative of a change in the 
characteristic host population. However, it is possible that 
deeper clustering observations at $z\sim1-2$ will 
eventually probe these luminosities, and test this prediction. 

Realistically, the luminosities of interest 
are sufficiently low that X-ray surveys present 
the most viable current probe, but with the small $\lesssim1\,{\rm deg}^{2}$ field 
sizes typical of most surveys, the quasar autocorrelation function cannot be 
constrained to the necessary accuracy to distinguish the models. However, 
as proposed in \citet{KH02}, 
the AGN-galaxy cross-correlation presents a possible solution. For example, 
there are sufficient galaxies at $z\sim1$ in the fields of surveys like 
e.g.\ DEEP2 or COMBO-17 (with field sizes $\sim3.5\,{\rm deg}^2$ 
and $\sim0.8\,{\rm deg}^{2}$, respectively)
that the accuracy of cross-correlation measurements 
is limited by the number of AGN; considering the 
hard X-ray selected AGN samples in the CDF-N or CDF-S (with field 
sizes $\sim0.01-0.5\,{\rm deg}^{2}$) from $z\sim0.8-1.6$ 
would represent a factor $\sim2-3$ increase in the number density of 
AGN over the \citet{Coil06.agn} sample in Figure~\ref{fig:bias.vs.lum}, while 
extending the AGN luminosities to a depth of $\sim3\times10^{9}\,L_{\sun}$ 
($\sim10^{42}\,{\rm erg\,s^{-1}\,cm^{-2}}$). The measurement 
of the cross-correlation between observed galaxies in these fields 
and deep X-ray selected faint AGN at $z\gtrsim0.5$ does, therefore, 
present a realistic means to test the differences in these models at low luminosities. 
The observation of a feature as shown in 
Figure~\ref{fig:bias.vs.lum} should correspond to a characteristic 
transition in the quasar host/fueling populations.

\section{Clustering of Different Populations}
\label{sec:populations}

\begin{figure*}
    \centering
    \figzoom
    \plotone{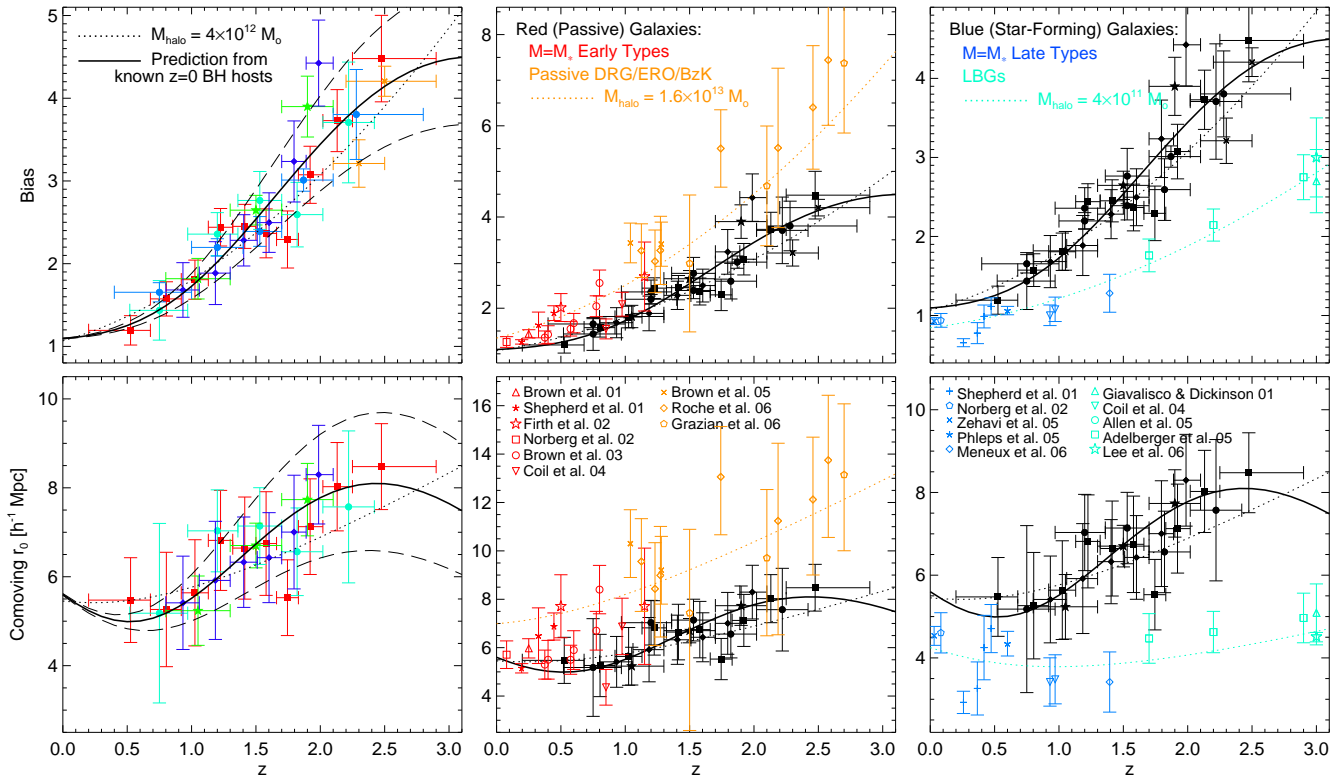}
    \caption{Clustering of quasars as a function of redshift
    (colored points in left panels, black in 
    other panels; as in Figure~\ref{fig:local.cluster.compare}), 
    compared to different models of possible host populations. Upper and lower 
    panels plot bias and comoving correlation length, respectively. 
    Solid line inverts the comparison in Figure~\ref{fig:local.cluster.compare}, 
    i.e.\ uses the estimated local clustering of red galaxies to predict 
    the quasar clustering assuming quasars are the progenitors of 
    present ellipticals (long dashed lines show $\sim1\,\sigma$ range 
    from uncertainties in local bias and observed bright quasar Eddington ratios). 
    Dotted lines correspond to halos of constant mass (as labeled). 
    Center panel compares this with the observed clustering of early-type 
    galaxies (at the characteristic red galaxy $M_{\ast}$ or $L_{\ast}$ at $z\lesssim1$; 
    at higher redshift $b(M_{\ast})$ is not longer well determined, so various 
    passive galaxy surveys are considered). Right panels compare 
    with late-type galaxies. 
    \label{fig:bias.populations}}
\end{figure*}

In Figure~\ref{fig:bias.populations} we compare the 
observed quasar bias and correlation length as a function of redshift 
with the expected clustering of quasar hosts, i.e.\ 
evolving the observed bias of BH (quasar ``relic'') hosts {\em up} from $z=0$.
A $z=0$ elliptical galaxy or spheroid of stellar mass $\mgal$ has 
a bias $b(M,\,z=0)$ shown in Figure~\ref{fig:local.cluster.compare}, 
and a BH of mass $M_{\rm BH}\approx0.001\,\mgal$.
For convenience, we adopt the analytic fit to $b(\mgal,\,z=0)$ in \citet{Li06}, 
\begin{equation}
b(\mgal,\,z=0)/b_{\ast}=0.896+0.097\,(\mgal/M_{\ast}),
\label{eqn:local.bias}
\end{equation}
where here 
$M_{\ast}=1.02\times10^{11}\,M_{\sun}$ is the Schechter function ``break'' mass
\citep{Bell03}
and $b_{\ast}=1.2$ for red galaxies \citep{Zehavi05,Li06}.
The progenitors of 
these systems therefore represent the ``characteristic'' active 
systems when the QLF characteristic luminosity $L_{\ast}(z)$ 
is given by $L_{\ast}(z)\approx \lambda\,L_{\rm Edd}(0.001\,\mgal)$, i.e.\ 
when these BHs dominated the $\sim L_{\ast}$ quasar population 
and assembled most of their mass. Evolving the 
local observed bias of the $z=0$ spheroids, $b(\mgal,\,z=0)$, to this redshift 
with Equation~(\ref{eqn:bias.evol}) yields the expected bias 
that the quasar hosts (and therefore quasars themselves) 
at this redshift should have, $b_{Q}(z)$. 
For future comparison, this is approximately given by 
\begin{equation}
b_{Q}(z)\approx 1+0.014\,D(z)^{-1}\,10^{5.70\,x-2.30\,x^{2}-3.35\,x^{3}}, 
\label{eqn:bestfit.bias}
\end{equation}
where $x\equiv \log{(1+z)}$ and $D(z)$ is again the linear growth factor. 
This expectation is plotted, with the $\sim1\sigma$ combined uncertainties 
from errors in the measured QLF $L_{\ast}$ and local 
bias $b(\mgal,\,z=0)$, comparable to the inherent factor $\lesssim2$ 
systematic uncertainty in the appropriate Eddington ratios 
and bolometric corrections. We also plot the corresponding 
correlation length $r_{0}$; because measurements of this quantity are 
covariant with the fitted correlation function slope $\gamma$, we 
renormalize the models and observations to 
\begin{equation}
r_{0}' \equiv 8\,h^{-1}\,{\rm Mpc}\,
{\Bigl (}\frac{r_{0,\,\rm fit}}{8\,h^{-1}\,{\rm Mpc}}{\Bigr )}^{{\gamma}/{1.8}}. 
\end{equation}
This is similar to the non-linear $\sigma_{8}^{\rm NL}$ parameter 
\citep[standard deviation of galaxy count fluctuations in a sphere of 
radius $8\,h^{-1}\,{\rm Mpc}$, i.e.\ $\sigma_{8}$ measured for 
an evolved density field, see][]{Peebles80}, and 
effectively compares the amplitude of clustering at $8\,h^{-1}\,{\rm Mpc}$ with a 
fiducial model with $\gamma=1.8$, minimizing the covariance.

\begin{figure}
    \centering
    \figzoom
    \plotter{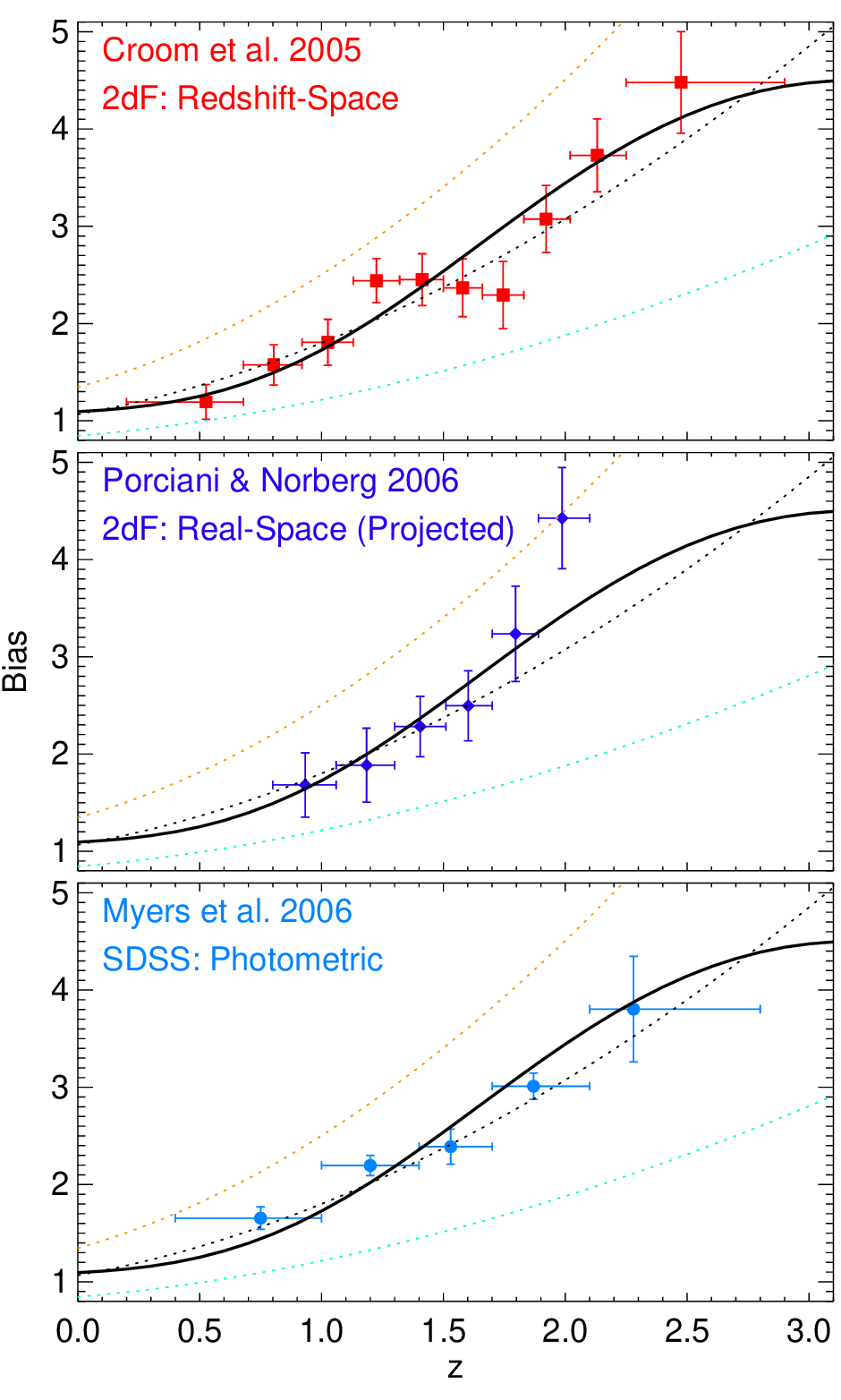}
    \caption{As Figure~\ref{fig:bias.populations} (upper left), but showing only single quasar 
    clustering measurements to highlight the significance of these comparisons from any 
    {\em individual} survey. The two 2dF results are not independent, but use different methods 
    to derive the quasar bias. 
    \label{fig:bias.populations.independent}}
\end{figure}

The expectation agrees well with observed 
quasar clustering as a function of redshift ($\chi^{2}/\nu=29.6/32$, with no free parameters). 
For comparison, we plot the expected clustering of halos of a 
fixed mass, $M_{\rm halo}\sim4\times10^{11}-10^{13}\,h^{-1}\,M_{\sun}$, 
determined in the context of linear collapse theory following \citet{MoWhite96}, 
modified according to \citet{ShethTormen01} and with the power spectrum calculated 
following \citet{EH99} for our adopted cosmology.
As noted in most previous studies \citep[e.g.,][]{Porciani04,Croom05}, a constant host halo mass 
of a few $10^{12}\,h^{-1}\,M_{\sun}$ provides a surprisingly good fit to the 
trend with redshift. This empirical fit, 
is comparable to our expectation from 
elliptical populations (best-fit halo mass $3.86\times 10^{12}\,h^{-1}\,M_{\sun}$
with $\chi^{2}/\nu=28.9/31$; of course, the 
exact best-fit mass depends systematically on cosmology, but this 
conclusion is robust). 
There is at most a marginal trend for the halo mass 
to increase with redshift (for $M_{\rm halo}\propto(1+z)^{k}$, the best-fit $k=0.41\pm0.45$; 
corresponding to a $\sim50\%$ increase over the observed redshift range). 

Note that the measurements 
shown are not all statistically independent, and the significance of this comparison 
will diminish if we consider any single quasar clustering measurement. 
Figure~\ref{fig:bias.populations.independent} demonstrates the same comparison, 
highlighting individual quasar bias measurements separately. However, the 
previous agreement and our conclusions 
are similar in all cases. As discussed by the authors, \citet{Porciani06} find 
a somewhat higher clustering amplitude in their highest-redshift ($z\sim2$) bin than 
\citet{Croom05} studying the same sample (and higher than \citet{Myers06} and 
\citet{Adelberger05.lifetimes} who study independent samples), but the 
significance of the \citet{Porciani06} result is $\lesssim2\,\sigma$. 

That a constant halo mass fits the data as well as observed 
suggests that there may be a physical driver or triggering mechanism associated 
with these halos. It is suggestive that this corresponds to the ``group scale;'' i.e.\ 
minimum halo mass of small galaxy groups, in which galaxy-galaxy mergers 
are expected to proceed most efficiently. However, the redshift evolution of 
this threshold is not well-determined \citep[but see][who find a similar 
``group scale'' halo mass at $z=1$]{Coil06.groups}, 
nor is the rate or behavior of merging within such 
halos. An a priori theoretical model for the prevalence of quasars in halos of 
this mass is therefore outside the scope of this paper, but remains an important topic for 
future work. 

Since it is also established, as discussed in \S~\ref{sec:local}, that the 
characteristic mass of active BHs increases with redshift, this implies substantial 
evolution in the ratio of BH to host halo mass to $z\sim2$. It is unclear how much of 
this may owe to evolution in the ratio of BH to host stellar mass: observational estimates 
imply some such evolution \citep[e.g.,][]{Shields05,Peng06,Woo06}, but 
upper limits from evolution in stellar mass densities 
\citep{H06.mbhmhost} allow only a factor $\sim2$ evolution 
by $z=2$. Therefore, there might also be at least some 
increase with redshift in the characteristic ratio of stellar to halo mass. Future 
constraints from halo occupation models or galaxy clustering at high redshifts will be 
valuable in breaking this degeneracy, and potentially provide important clues to galaxy 
assembly histories. 

We contrast these predictions with two extremely simple models. In the first, quasar 
activity is an unbiased tracer of dark matter, i.e.\ $b(z)=1$. This does, after all, 
appear true at $z=0$ \citep{Wake04,Grazian04,Constantin06}. This 
is immediately strongly ruled out: there is an unambiguous 
trend that higher-redshift quasars are more strongly biased (as 
noted in essentially all observed quasar correlation functions). 

Next, we consider the possibility that quasars live in the {\em same} 
halos at all redshifts. This is equivalent to some ``classical'' 
interpretations of pure luminosity evolution in the QLF  \citep[e.g.,][]{Mathez76}; i.e.\ that 
quasars are cosmologically-long lived
\citep[although other observations demand a lifetime $\lesssim10^{7}$\,yrs; e.g.]
[and references therein]{Martini04}, and dim from 
$z=2$ to the present. 
It is also equivalent to saying that quasars are triggered, even for a
short time, in the same objects over time (e.g.,\ stochastically or 
by some cyclic mechanism). 
In this case, the quasar lifetime can still 
be short (with a low ``duty cycle'' $\delta\sim10^{-3}$), although 
Eddington ratios must still tend to increase with redshift. 
Then, the halo bias evolves as Equation~(\ref{eqn:bias.evol}), 
from a $z=0$ value $b(0)\sim1.0-1.2$ \citep[][]{Norberg02}. 
Although this model is qualitatively consistent with some quasar observations, 
it is not nearly sufficient to explain the evolution of clustering amplitude with redshift and 
is ruled out at very high ($>10\,\sigma$) significance. As noted 
in previous studies of quasar clustering \citep[][]{Croom05}, 
quasars at different redshifts must reside in {\em different} parent halo populations; 
quasars cannot, as a rule, be long lived or recurring/episodic/cyclic 
(although this does {\em not} apply to very low-accretion rate activity, perhaps 
associated with ``radio modes''; see e.g.\ Hopkins, Hernquist, \& Narayan 2005). 

Rather than a uniform population of halos at all redshifts, 
what if quasars uniformly sample observed galaxy populations? 
It is, for example, easy to modify the above scenario slightly: quasars 
are cosmologically long-lived or uniformly cyclic/episodic, but only 
represent the present/extant population of BHs (equivalently, the 
present population of spheroids). In this case, 
quasar correlation functions should uniformly trace early-type 
correlations at all redshifts. 

Figure~\ref{fig:bias.populations} compares 
observed early-type/red galaxy clustering as a function of redshift 
with that measured for quasar populations. At low redshifts $z\lesssim1$, 
both mass functions and clustering as a function of mass/luminosity  
are reasonably well-determined, so we plot clustering at the characteristic 
early-type (Schechter function) $M_{\ast}$ or $L_{\ast}$. At higher redshift, 
caution should be used, since this characteristic mass/luminosity is 
not well determined, and so we can only plot clustering of observed massive 
red galaxies (which, given the observed dependence of clustering 
amplitude on mass/luminosity and color, may bias these estimates to high 
$b(z)$ if surveys are not sufficiently deep to resolve $M_{\ast}$ or $L_{\ast}$). 
There is also the additional possibility that the poorly known redshift distribution 
of these objects may introduce artificial scatter in their clustering estimates. 
Bearing these caveats in mind, the clustering of quasars and red galaxies are inconsistent at 
high ($>6\sigma$) significance. 
Quasars do {\em not} uniformly 
trace the populations of spheroids/BHs which are ``in place'' 
at a given redshift. Note, however, that in this comparison the 
systematic errors almost certainly dominate the formal statistical uncertainties, so 
the real significance may be considerably lower.

An alternative possibility is that BH growth might uniformly 
trace star formation. In this case, quasar clustering 
should trace the star-forming galaxy population. 
Figure~\ref{fig:bias.populations} compares 
observed late-type/blue/star-forming galaxy clustering as a function of redshift 
with that observed for quasar populations. Again, at $z\lesssim1$ we 
plot clustering at the characteristic 
$M_{\ast}$ or $L_{\ast}$. At higher redshift 
we can only plot ``combined'' clustering of observed 
star-forming populations (generally selected as Lyman-break galaxies); 
again caution is warranted given the known dependence of 
clustering on galaxy mass/luminosity \citep[for LBGs, see][]{Allen05}. 
In any case, the clustering is again inconsistent at 
high ($>10\sigma$) significance. Quasars do {\em not} uniformly 
trace star-forming galaxies. This appears to be contrary to some previous claims 
\citep[e.g.,][]{Adelberger05.lifetimes}; however, in most cases where 
quasars have been seen to cluster similarly to blue galaxies, either 
{\em faint} AGN populations (not $\sim L_{\ast}$ quasars) or 
bright ($\gg L_{\ast}$) blue galaxies were considered. Indeed, quasars 
do cluster in a manner similar to the {\em brightest} blue galaxies 
observed at several redshifts \citep[e.g.,][at $z\sim1$ and 
$z\gtrsim2$, respectively]{Coil06.lumdept,Allen05}. This should not be surprising; 
since quasars require some cold gas supply for their fueling, they cannot be significantly 
more clustered than the most highly clustered (most luminous) population of 
galaxies with that cold gas. Again, this highlights the fact that the real systematic 
issues in this comparison probably make the significance considerably less than the 
formal $\sim10\,\sigma$ seen here.  

\begin{figure}
    \centering
    \figzoom
    \plotone{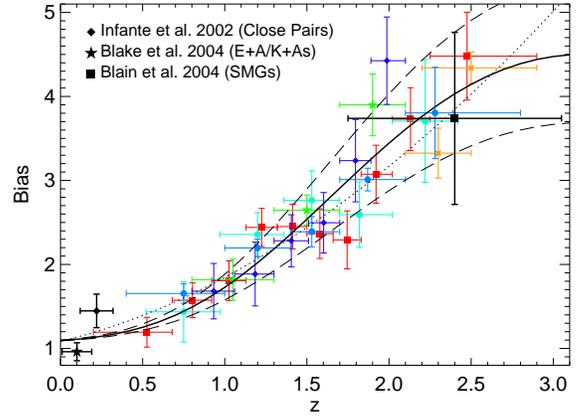}
    \caption{As Figure~\ref{fig:bias.populations}, but comparing observed quasar 
    clustering (colored points) 
    as a function of redshift to that various populations usually associated with 
    galaxy mergers (black points): post-starburst (E+A/K+A) galaxies, 
    close galaxy pairs, and sub-millimeter galaxies. 
    \label{fig:compare.smg}}
\end{figure}

We would also like to compare quasar clustering directly to the 
clustering of gas-rich (luminous) mergers. Figure~\ref{fig:compare.smg} 
attempts to do so, using available clustering measurements for 
likely major-merger populations. At low redshifts, \citet{Blake04} have 
measured the clustering of a large, uniformly selected 
sample of post-starburst (E+A or K+A) elliptical galaxies in the SDSS, 
which from their colors, structural properties, and fading morphological 
disturbances \citep[e.g.,][and references therein]{Goto05} are believed to 
be recent major merger remnants. \citet{Infante02} have also measured the 
large-scale clustering of close galaxy pairs selected from the SDSS at 
low redshift. At high redshift, no such samples exist, but \citet{Blain04} 
have estimated the clustering of a moderately large sample of 
spectroscopically identified sub-millimeter galaxies at $z\sim2-3$, 
for which the similarity to local ULIRGs in 
high star formation rates, dust content, line profiles, and disturbed morphologies 
suggests they are systems undergoing major, gas-rich mergers 
\citep[e.g.,][and references therein]{Pope05,Pope06,Chak06}. The clustering of these 
populations is consistent at each redshift with observed quasar clustering
\citep[see also][]{H06d}. 
This is contrary to the conclusions of \citet{Blain04}, but they compared 
their SMG clustering measurement with earlier quasar clustering 
data \citep{Croom02} below their 
median redshift $z\sim2.5$. Figure~\ref{fig:compare.smg} 
demonstrates that the dependence of quasar 
clustering on redshift is such that at the same redshifts, the two agree very well. 
However, given the very limited nature of the data, and the lack of 
a uniform selection criteria for ongoing or recent mergers at different 
redshifts, we cannot draw any strong conclusions from the direct 
merger-quasar clustering comparison alone.

Although quasars do not appear to trace star-forming galaxies, 
\citet[][and references therein]{Adelberger05.lbgclustering} 
have shown that the star formation rates, clustering properties, and number densities 
of high-redshift LBGs suggest they are the progenitors of present-day ellipticals. 
To the extent that quasars are also the progenitors of ellipticals (but with a larger 
clustering amplitude at a given redshift compared to LBGs), this suggests a crude
``straw-man'' outline of an 
evolutionary sequence with time, from LBG to quasar to remnant elliptical galaxy. 
Knowing how the clustering properties of halos hosting LBGs with a given observed bias 
at some redshift $z_{\rm LBG}$ will subsequently evolve, we can determine the 
redshift $z_{Q}$ at which this matches observed quasar clustering properties. This 
offset, if LBGs and quasars are indeed subsequent progenitor ``phases'' in the sequence of 
evolution to present day ellipticals, defines the ``duration'' of the LBG ``phase'' or time
between LBG and quasar ``stages.'' 

\begin{figure}
    \centering
    \figzoom
    \plotter{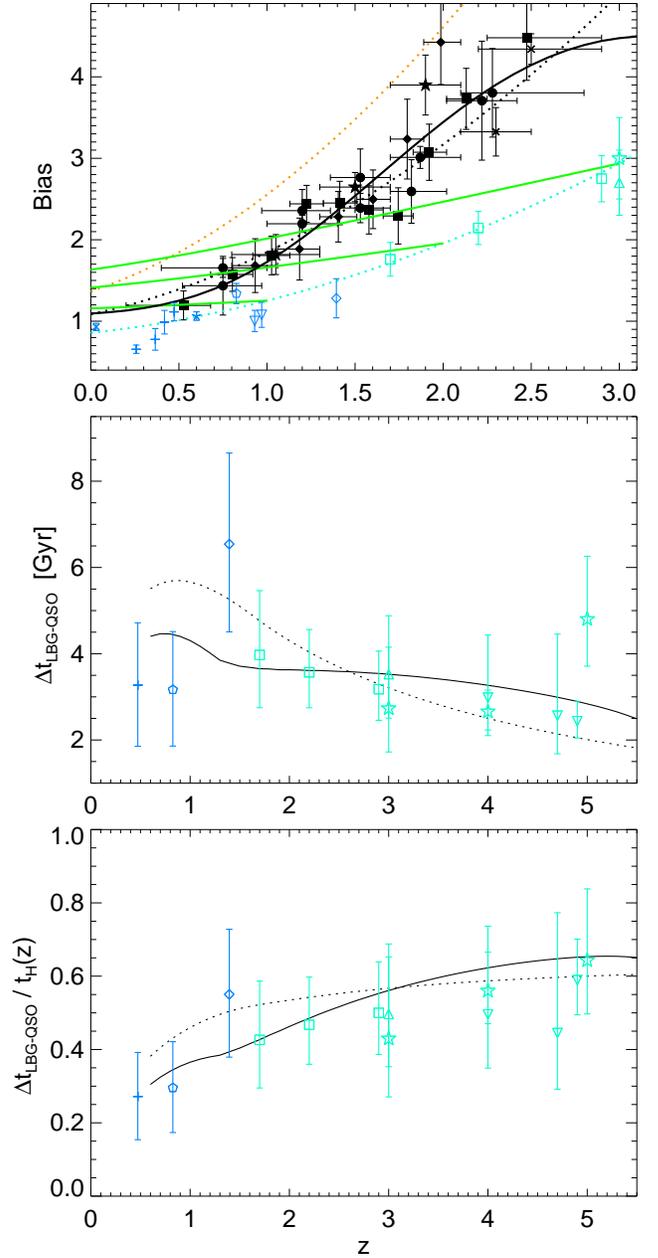}
    \caption{{\em Upper:} Clustering of quasars and star-forming galaxies, 
    as in Figure~\ref{fig:bias.populations}. Solid green lines show the subsequent evolution 
    of the clustering of the star-forming galaxy halos from $z=1,\,2,\,3$. 
    {\em Middle:} Time delay between the star-forming or LBG phase and ``quasar'' epoch, 
    defined as the time after the observed redshift of each LBG population at which 
    its evolved clustering will match that of the observed quasar population. Points as in 
    upper panel, with data from \citet[][cyan inverted triangles]{Ouchi04} added at $z>4$. Dashed 
    and dotted lines show time for halos of mass $\sim4\times10^{11}\,h^{-1}\,M_{\sun}$ (the 
    typical LBG host mass) at each redshift to reach $\sim4\times10^{12}\,h^{-1}\,M_{\sun}$ 
    (dotted) or the (weakly redshift-dependent) halo mass defined by our best-fit trend 
    in the upper panel (solid). 
    {\em Lower:} As the middle panel, but the time shown is as a fraction of the 
    Hubble time at the ``quasar'' epoch.
    \label{fig:lbg.ages}}
\end{figure}

Figure~\ref{fig:lbg.ages} considers this in 
detail. We show the observed clustering of quasars and LBGs from Figure~\ref{fig:bias.populations}, 
with curves illustrating the subsequent clustering evolution of the LBG 
host halos observed at $z=1$, $z=2$, and $z=3$. These correspond to the characteristic 
observed quasar clustering at $z=0.4$, $z=1.0$, and $z=1.3$, respectively. 
Thus, halos of the characteristic 
LBG host halo mass at $z=3$ will grow to the characteristic quasar host mass at $z=1.3$, 
and so on. We also show the physical time corresponding to this offset, calculated from 
the observed LBG clustering at various redshifts and the best-fit estimate of 
the LBG host mass $\sim4\times10^{11}\,h^{-1}\,M_{\sun}$, and this time divided by the 
Hubble time (age of the Universe) at the ``quasar epoch'' $z_{Q}$. Interestingly, this implies 
that objects characteristically spend $\sim3-4$\,Gyr ($\sim1/2\,t_{H}$ at the redshifts of interest) 
in the ``LBG phase.'' This may reflect the time for dark matter halos to grow from the 
characteristic LBG mass, at which 
star formation and the conversion of mass to light appears to be most efficient 
\citep[e.g.,][]{WhiteRees78} to the typical quasar host mass; but it is 
also possible that associated physical 
processes related to quasar fueling or the termination of star formation set this timescale.
If quasars are triggered in major mergers, this rather large time offset 
\citep[as opposed to the typical $\sim100\,$Myr delay between starburst and quasar 
in major merger simulations,][]{DiMatteo05} implies that LBGs are themselves 
not primarily driven in major mergers. A similar conclusion was recently reached by 
\citet{Law06} from direct analysis of LBG morphologies at $z\sim2-3$.
This conclusion and the LBG clustering 
in Figure~\ref{fig:lbg.ages} \citep{Wechsler01} are broadly consistent with the expectations 
of semi-analytic models \citep{Somerville01} which argue LBGs are driven largely 
by ``collisional'' minor merging. 

\begin{figure}
    \centering
    \figzoom
    \plotter{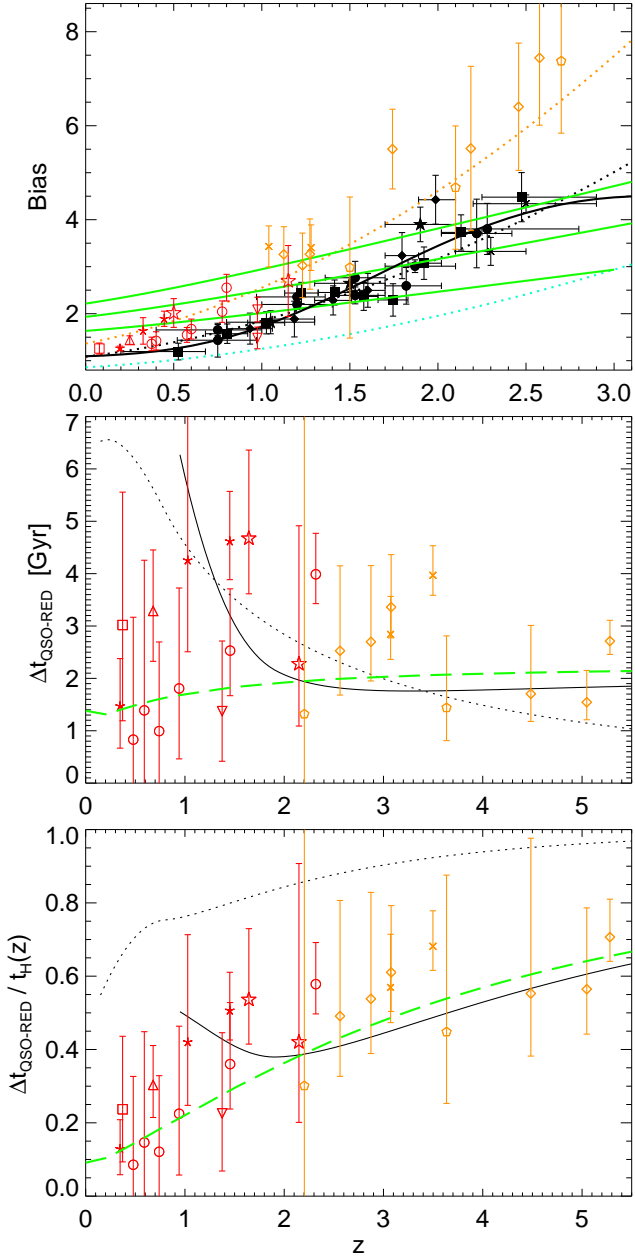}
    \caption{As Figure~\ref{fig:lbg.ages}, but instead showing the time from the ``quasar phase'' 
    to the ``red galaxy'' phase implied by the observed clustering of both populations. 
    Green long-dashed lines in the middle and lower panels 
    show the time required for the ``burst-quenched'' star formation history models 
    from \citet{Harker06} (which yield a good empirical approximation to the buildup and mean 
    color evolution of red galaxies) to redden to a threshold $U-B>0.35$. 
    \label{fig:red.ages}}
\end{figure}

We can also use this approach to determine the time between 
the ``quasar'' and red/elliptical phases
in this evolutionary sequence. Figure~\ref{fig:red.ages} 
shows this, in the style of Figure~\ref{fig:lbg.ages}, where 
the redshift shown in the middle and lower panels refers to 
the redshift of the observed quasar population, and the time to the delay at which their evolved clustering matches that measured for the red galaxy population. Note that the continuous curves 
calculated in the middle and lower panels assume the red galaxy clustering is well-fitted 
by the plotted (upper panel) constant halo mass $\sim1.6\times10^{13}\,h^{-1}\,M_{\sun}$ curve; 
this is, in fact, not a very good approximation at low redshifts, hence these curves diverge 
below $z\sim1-2$ from the times calculated from the actual red galaxy clustering measurements. 

In the lower panels, we also plot the time for ``burst-quenched'' star formation 
history models adapted from \citet{Harker06} to redden to a typical constant ``red galaxy'' threshold 
rest-frame color $U-B>0.35$. These model star formation histories assume a constant 
star formation rate until $1\,$Gyr before the ``quasar epoch,'' then a factor $5$ enhanced 
star formation rate for this $1\,$Gyr, at which point star formation ceases. Essentially, this 
yields a useful toy model for ``quenching,'' if indeed the triggering of quasars is associated 
with the formation of ellipticals or termination of star formation (the pre-quenching 
enhancement being an approximation to, e.g., merger-induced star formation enhancements), 
which \citet{Harker06} 
demonstrate yields a reasonable approximation to the observed mean color, 
number density, and Balmer H$\delta_{\rm F}$ absorption strength evolution of red galaxies. 
The predicted time for such quenched star formation histories to redden to typical 
red galaxy colors agrees well with the time estimated from clustering here at all redshifts;
i.e.\ the color and halo mass evolution of these systems are consistent with 
reasonable star formation histories in which quasar activity is associated with ``quenching'' 
or the termination of star formation.

We have estimated the time offsets in Figures~\ref{fig:lbg.ages} \&\ \ref{fig:red.ages} 
from a direct comparison of the observed clustering. Instead,  
one might imagine adopting the implied halo mass ($\sim4\times10^{11}\,h^{-1}\,M_{\sun}$)
at the ``star-forming'' phase 
and using extended Press-Schechter (EPS) theory to calculate 
the average time for a typical progenitor halo of this mass at each observed 
redshift to grow to the implied quasar host halo mass 
($\sim4\times10^{12}\,h^{-1}\,M_{\sun}$).  We discuss 
this in greater detail in \S~\ref{sec:ages}, and show that it has no effect on 
our conclusions. For the purposes here, 
adopting this methodology (specifically, calculating the evolution of the 
``main branch'' progenitor halo mass with redshift following \citet{Neistein06} in 
our adopted cosmology) systematically increases the inferred time delays (points)
in Figure~\ref{fig:lbg.ages} by $\sim1-2$\,Gyr and those in Figure~\ref{fig:red.ages} 
by $\sim0.5-1$\,Gyr, but does not significantly change the 
plotted trends or comparisons. 

So, this leaves us with the following suggested empirical picture of galaxy evolution. 
Galaxies form and experience a typical ``star forming'' or LBG epoch, with maximal 
efficiency around a characteristic halo mass of a few $\sim10^{11}\,h^{-1}\,M_{\sun}$. 
Growth continues, presumably via normal accretion, minor mergers, and star 
formation, for roughly half a Hubble time, until systems have growth to a 
characteristic halo mass $\sim4\times10^{12}\,h^{-1}\,M_{\sun}$. At this point, 
some mechanism (for example, a major merger, as this may be the characteristic scale 
at which the host halo grows large enough to host multiple ``large'' star-forming systems) 
triggers a short-lived ``quasar'' phase, drives a morphological transformation from 
disk to spheroid, and terminates star formation. About $\sim1-2$\,Gyr after this, the host 
halos have grown to $\sim10^{13}\,h^{-1}\,M_{\sun}$ and the spheroids 
have reddened sufficiently to join the typical ``red galaxy'' population. They 
then passively evolve (although they may experience some gas-poor or ``dry'' 
mergers) to $z=0$, satisfying observed correlations between BH and spheroid properties. 
Although individual BHs can, in principle, gain significant mass from 
``dry'' mergers \citep[see, e.g.][]{Malbon06}, 
this cannot (by definition) add to the total mass budget of BHs, which 
must be built up via accretion. 
Note that this is only a rough conception outline of an ``average'' across 
populations and should not be taken 
too literally. Different systems will undergo these processes at different times, 
and (possibly) via different mechanisms. Still, this provides a potentially 
useful framework in which to interpret these observations. 

\section{Age-Mass Relations and Clustering}
\label{sec:ages}

\begin{figure*}
    \centering
    \figzoom
    \plotone{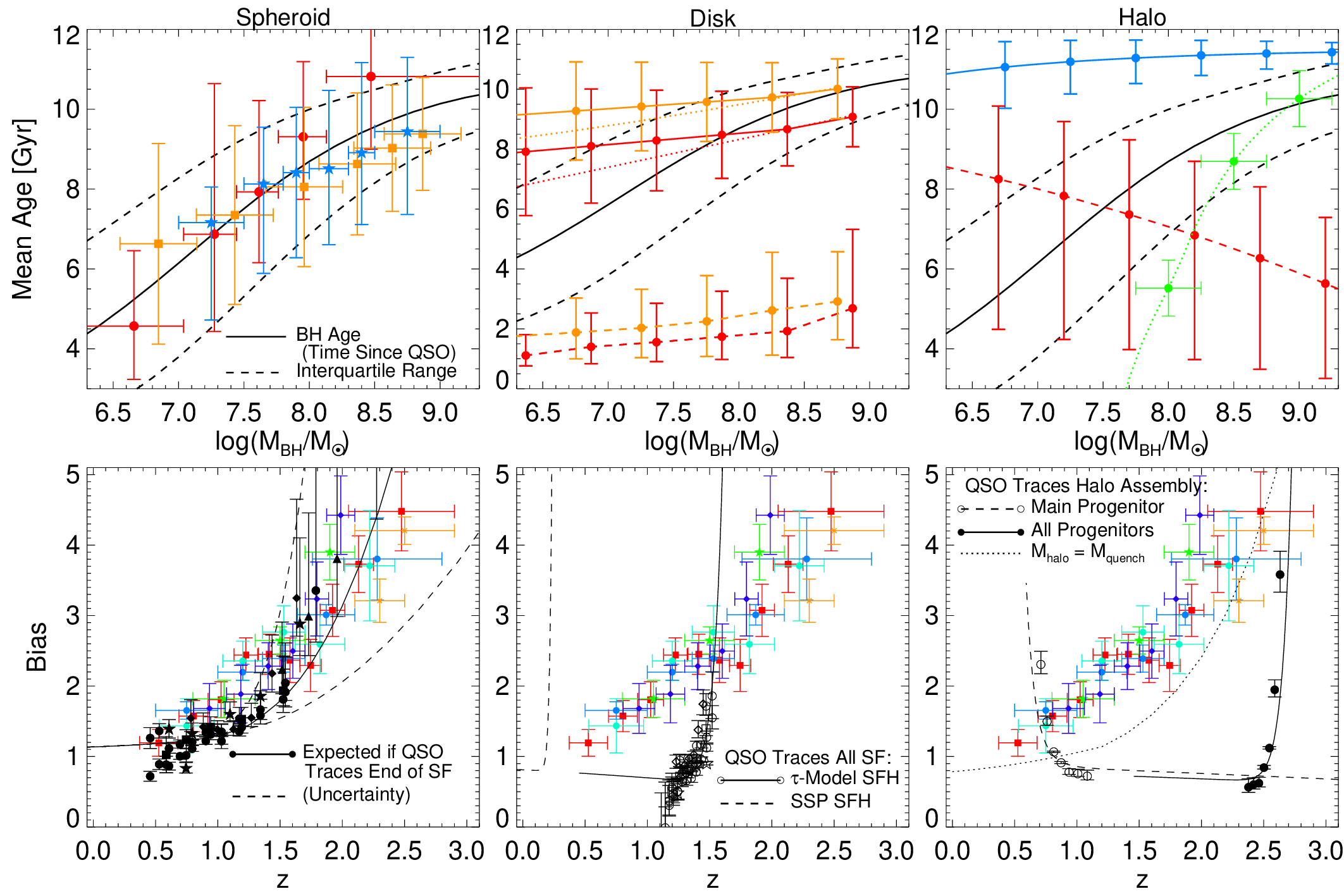}
    \caption{
    {\em Left:} Upper panel shows the mean $z=0$ age (lookback time to the mean ``formation epoch'') 
    of BHs as a function of mass (black solid line, dashed lines show $25\%-75\%$
    quartile ranges), compared to the stellar population age of their hosts. 
    Ages of spheroids as a function of mass (with $\mbh=\mu\,\mgal$) are shown (colored points) from 
    \citet[][NFPS; red squares]{Nelan05}, \citet[][SDSS; blue stars]{Gallazzi06}, \citet[][orange circles; 
    ``field'' subsample]{Thomas05}. Errors show 
    the {\em dispersion} in ages at a given mass. 
    Lower panel uses this age to predict quasar clustering as a function of redshift; i.e.\ assuming 
    the ``quasar epoch'' of spheroids of a given mass is associated with the 
    termination of star formation (black lines, as labeled; colored points show observed 
    quasar clustering as in Figure~\ref{fig:bias.populations}). 
    {\em Center:} Same, for ages of host disks; ages from $\tau$-model fits of 
    \citet[][red]{BelldeJong00} and \citet[][orange]{Gavazzi02} 
    (the offset between them owes to the choice of initial time in the $\tau$-model). 
    Solid lines asumme $M_{\rm BH} \propto M_{\rm bulge}$, 
    dotted $M_{\rm BH}\propto (M_{\rm disk}+M_{\rm bulge})$. Dashed lines re-calculate the 
    age for a single-burst SFH. Lower panel is as lower left, assuming quasar activity is 
    associated with the star formation epoch (as labeled). 
    {\em Right:} Blue (solid) line 
    shows the ``all progenitor'' age \citep[DM ``downsizing'' from][]{Neistein06}, 
    red (dashed) the age of the main progenitor halo, and 
    green (dotted) the time when halo crossed the ``quenching'' mass from 
    \citet{Dekel06}. Lower panel as lower left, assuming quasar age is equal 
    to the halo age as labeled. 
    \label{fig:age.bias.compare}}
\end{figure*}

In Figure~\ref{fig:age.bias.compare}, we compare the mean age of BHs of a 
given $z=0$ mass with that of the stellar population of their hosts. 
At a given redshift, the characteristic QLF luminosity $L_{\ast}(z)$ and 
corresponding characteristic ``active'' mass $M_{\rm BH}$ from 
Figure~\ref{fig:active.mbh}
define the epoch of growth of BHs of that mass. The typical 
``age'' of BHs of that mass will be the time since this 
epoch. In detail, Equation~(\ref{eqn:model}) relates the observed QLF 
to the rate at which BHs of a given relic mass are formed as a function of 
redshift. We adopt the fits to this equation given in \citet{HRH06}, which 
use the model quasar lightcurves determined in \citet{H06a,H06b} to 
calculate the time-averaged rate of formation of individual BHs as a function of 
mass and redshift, to estimate the median age (peak in rate of formation/creation of 
such BHs as a function of time) and $25-75\%$ interquartile range in 
``formation times.'' This introduces some model dependence, but as 
discussed in \S~\ref{sec:local} a similar result is obtained using 
very different methodologies, including 
purely empirical, simplified models \citep{YT02,Merloni04.bhmf,Marconi04,Shankar04}, or 
direct calculation from 
observed Eddington ratio distributions \citep{Vestergaard04,MD04,Kollmeier05}.

In any case, we recover the well-known trend that the more massive BHs 
are formed at characteristically earlier times \citep{Salucci99,YT02,Ueda03,Heckman04,HMS05,
Merloni04.bhmf,Marconi04,Shankar04,MD04,Kollmeier05}. This is not surprising, as most 
massive BHs must be in place by $z\sim2$ to power the brightest quasars, 
and these objects are generally ``dead'' by low redshift 
\citep[with lower-mass objects dominating the local QLF, e.g.][]{Heckman04}.

Given a BH mass, we can compare with the observationally 
determined age of the typical host galaxy (with $\mbh=\mu\,\mgal$). 
First, we consider early-type hosts, specifically the stellar 
ages of host spheroids of BHs at each mass $\mbh$. 
The mean ages (and dispersion about that mean) of ellipticals as a function 
of stellar mass have been estimated in a number of studies, 
recently for example by \citet{Gallazzi05,Gallazzi06}, 
who fit SDSS spectra (line indices) and photometry for $\sim175,000$ local 
galaxies to various realistic star formation histories, including a 
mix of continuous and/or starbursting histories while allowing mass, total 
metallicity, and abundance ratios to freely vary. They quote 
$r$-band light-weighted ages, which for our purposes are effectively 
equivalent to the ages determined by fitting a single stellar population (SSP) 
or ``single burst'' model to observed spectra, and indeed agree very well 
with best-fit SSP ages from similar SDSS samples \citep{Clemens06,Bernardi06}
and previous studies 
\citep[][for a review see Renzini 2006]{Jorgensen99,Trager00,Kuntschner01,
Caldwell03,Bernardi05,Nelan05,Thomas05,Collobert06} 
as a function of elliptical stellar mass. A similar result is also obtained 
independently by \citet{Treu05} and \citet{diSeregoAlighieri06} 
from studies of the the fundamental plane 
evolution of early-type galaxies. Note that the error bars shown are 
the measured dispersions in the population about the mean age at a given mass, 
not the uncertainties in the mean ages themselves \citep[which are smaller; $\sim0.2$\,Gyr
statistical, $\sim1$\,Gyr systematic; see][]{Nelan05}. 
The agreement between BH and host stellar ages 
is good at all masses; both the trend and dispersion (interquartile or 
$\pm1\sigma$ range) about it are similar ($\chi^{2}/\nu\sim8/17$ for a direct comparison). 

If the age of its stellar populations is indicative of when the ``quasar epoch'' 
occurred in a given host, then, 
without making {\em any} assumptions about the masses of black holes or 
quasar Eddington ratios, we can use the mean age of stellar populations 
to predict quasar clustering. In this scenario, ellipticals of mass $M$, with 
mean age $t_{\rm host}$, would represent the population ``lighting up'' as 
quasars at a lookback time of $t_{\rm host}$, and so the 
quasar bias at that lookback time should be the local 
bias of ellipticals of mass $M$ (Equation~\ref{eqn:local.bias}), 
evolved to the appropriate lookback time $t_{\rm host}$ 
with Equation~(\ref{eqn:bias.evol}). Figure~\ref{fig:age.bias.compare} 
compares this expectation with the observed quasar bias as a function of 
redshift. Despite the very simple nature of this model, which 
ignores both the range of ages at a given $M$ and, similarly, the range in 
host masses at a given time, the agreement is reasonable. 
Including the dispersion in ages, i.e.\ modeling the age distribution at each 
$\mbh$ as a Gaussian with the observed scatter, improves the 
agreement and yields a nearly identical prediction of bias as a function 
of redshift to that in Figure~\ref{fig:bias.populations} (solid black line).

We can of course repeat these exercises for 
other possible ``host'' populations. 
We next consider correlations with the star formation 
histories of late-type or star-forming host galaxies -- i.e.\ 
the possibility that quasar activity 
is generically associated with star formation.
The observed star formation histories are similarly estimated, generally by fitting 
to exponentially declining models ($\tau$-models; 
star formation rate $\dot{M}\propto\exp{[-(t-t_{i})/\tau]}$ since 
an initial cosmic time $t_{i}$). Specifically, we consider 
the fits of late-type ages as a function of stellar mass 
from \citet{BelldeJong00} and \citet{Gavazzi02} \citep[consistent with][]{Jansen00,Bell00,Boselli01,Kauffmann03.sfhs,PerezGonzalez03,
Brinchmann04,MacArthur04,Gallazzi05}. 
The mass-weighted age is calculated from the model SFR 
\citep[see][Equation~3]{BelldeJong00}. 
\citet{BelldeJong00} and \citet{Gavazzi02} technically quote the 
age and metallicity as a function of $K$ and $H$ band absolute magnitudes, 
respectively, but given their quoted best-fit stellar population models 
at each luminosity, it is straightforward to 
calculate the corresponding mass-to-light ratios ($M/L_{K}$ and $M/L_{H}$) 
and convert the observed luminosities to total stellar masses. 
To convert to a corresponding BH mass, we consider first 
a uniform application of the local BH-host mass relation, i.e.\ assuming 
BH mass is correlated with {\em total} stellar mass, and second 
determining the mean bulge-to-disk ratio for a given total late-type stellar 
mass or luminosity \citep[see][for the appropriate mean 
$B/T$ for different masses/luminosities]{Fukugita98,AllerRichstone02,Hunt04} 
and assuming BH mass is correlated with the bulge mass only.
Because the trend of age as a function of stellar mass is weak, 
considering the total mass or bulge mass makes little difference in our comparison, 
and we subsequently consider the observationally preferred correlation 
between BH and bulge mass. 

The mean age of a given population 
derived from different model star formation histories will, of course, be weighted 
differently. To show the systematic effects of such a choice, we 
roughly estimate the equivalent age from a single burst or SSP model. 
We calculate the $z=0$ observed $(B-V)$ color at each mass from the 
mean best-fit $\tau$ models, and then calculate 
the corresponding age for the same $(B-V)$ of a single burst model (of the same 
metallicity as a function of mass) from the models of \citet{BC03} with a 
\citet{Salpeter55} IMF. Although most of the observations above find 
this SSP approximation is not good for star-forming galaxies, 
it illustrates an important point. The SSP ages are weighted towards the 
youngest, bluest stellar populations -- essentially functioning as 
an indicator of the most recent epoch of significant star formation, and 
are therefore quite young \citep[$\lesssim2\,$Gyr; similar to the typical 
time since recent low-level starbursting activity found in 
late-types with the more realistic star formation
histories in][]{Kauffmann03.sfhs}. 
However, the trend as a function of mass is unchanged and the 
overall agreement is worse. Therefore, while the systematic 
effects here are substantially larger than the measurement 
errors in mean age as a 
function of mass, they cannot remedy the poor agreement with the 
ages of BH populations. 

We again use this age as a function of total/bulge mass, 
and the observed clustering of late-type galaxies from Figure~\ref{fig:local.cluster.compare} 
at $z=0$, to estimate what the quasar bias as a function of redshift should be, 
if these systems were the hosts of quasars and their quasar epoch were 
associated with the age of their stellar populations. 
The predictions are inconsistent with the observations at high significance ($> 4.5\sigma$), 
regardless of the exact age adopted ($\tau$-model or SSP). 
In fact, the predicted clustering as a function of 
redshift is highly unphysical (owing 
to the fact that there is relatively little difference in ages, 
but strong difference in clustering amplitudes from the least to most massive disks). 
Ultimately, this demonstrates that the hypothesis that quasar activity 
{\em generically} traces star formation is unphysical. This is also supported 
by the fact that the integrated global star formation rate 
and quasar luminosity density evolve in a similar, but not identical 
manner from $z\sim0-6$ \citep[e.g.,][]{Merloni04.hosts}. 

We next consider the the possibility 
that quasar activity traces pure dark matter assembly processes -- i.e.\ that 
the buildup of BHs in quasar phases purely traces the formation of their host halos.  
Given the local BH-host stellar mass relation from \citet{MarconiHunt03}, and 
the typical halo mass as a function of early-type hosted galaxy mass 
calibrated from weak lensing studies by \citet{Mandelbaum06}, 
we obtain the mean host halo mass as a function of BH mass 
(mean $M_{\rm halo}\sim4\times10^{4}\,M_{\rm BH}$; 
although the relation is weakly nonlinear). For our 
adopted cosmology, we then calculate the mean age 
(defined as the time at which half the present mass 
is assembled) of the main 
progenitor halo for $z=0$ halos of this mass. Error bars are taken from 
an ensemble of random EPS merger trees following 
\citet{Neistein06}. Knowing the mass of a halo at a given redshift, we 
calculate its clustering following \citet{MoWhite96} as in \S~\ref{sec:populations}, 
and use this combined with the mean ages to estimate the expected 
quasar clustering as a function of redshift if quasars were associated 
with this formation/assembly of the main progenitor halo.
Although the exact age will depend on 
cosmology and the adopted ``threshold'' mass fraction at which 
we define halo ``age,'' the result is the same, namely we 
recover the well-known hierarchical trend in which the most massive objects are 
youngest, in contradiction with quasar/BH ages. 

However, \citet{Neistein06} have pointed out that the mean assembly time, 
considering {\em all} progenitor halos, can exhibit so-called ``downsizing'' 
behavior. We therefore follow their calculation of the mean age of all progenitors 
as a function of $z=0$ halo (and corresponding BH) mass, and 
also use this to estimate quasar clustering as a function of redshift.
Again, although the systematic normalization depends somewhat on our 
definitions, it is clear that the recovered ``downsizing'' trend is, 
as the authors note, a subtle effect, and not nearly strong enough 
(inconsistent at $>10\sigma$) to explain the downsizing of 
BH growth. Again, the absolute value of the age obtained can be systematically 
shifted by changing our definition of halo ``assembly time,'' but the trend 
is not changed and significance of the disagreement with BH formation times 
is still high. 

Certain feedback-regulated models predict that black hole mass 
should be correlated with halo circular velocity 
\citep[as $M_{\rm BH}\propto v_{c}^{5}$ or $\propto v_{c}^{4}$;][]{SR98,WyitheLoeb03}, rather than 
halo mass. To consider this, we have re-calculated the ``all progenitor'' and 
``main progenitor'' ages, but instead adopted the time at which the 
appropriate power of the circular velocity ($v_{c}^{5}$ or $v_{c}^{4}$) 
reaches half the $z=0$ value. Because, for a given 
halo mass, $v_{c}$ is larger at higher redshift, this 
systematically shifts both ages to higher values, but the trends are similar. 
In each case the resulting ages disagree with the quasar/BH ages at 
even higher significance. 

Alternatively, some models \citep[][]{Birnboim03,Binney04,Keres05,Dekel06} suggest that a qualitative 
change in halo properties occurs at a characteristic mass, 
above which gas is shock-heated and cannot cool efficiently, forming a 
quasi-static ``hot accretion'' mode in which quasar feedback can act efficiently. 
Although there is no necessary reason why quasar activity should be 
triggered by such a transition, its posited association with quasar 
feedback leads us to consider this possibility. 

Knowing the mean $z=0$ halo 
mass for a given $M_{\rm BH}$, we plot the age at which the main progenitor 
halo mass surpassed the critical ``quenching'' mass defined as a function of 
redshift in \citet{Dekel06}. 
Since this amounts to a nearly constant characteristic halo mass 
$\sim\times10^{11}-10^{12}\,M_{\sun}$, the expected quasar clustering as a function 
of redshift is not unreasonable (see Figure~\ref{fig:bias.populations}; 
there is a systematic offset, but this is sensitive to the adopted cosmology). 
However, this model actually predicts too steep a trend of age 
with mass (inconsistent at $>6\sigma$). The ages of the most massive systems are reasonable 
\citep[which, in comparison to 
the ages of ellipticals, has been widely discussed;][]{Dekel06,Croton05,deLucia06}, 
but the host halo of a typical $\sim10^{8}\,M_{\sun}$ BH (i.e.\ Schechter 
$M_{\ast}$) crosses the threshold halo mass a mere $\sim5$\,Gyr ago, 
predicting, in this simple model, that these BHs should have been the characteristic active systems  
at $z\sim0.5$ instead of the observed $z\sim1.2-1.5$. At lower masses, 
the mean $z=0$ halos are only just at, or are still below, this critical halo mass. 
Given the scatter in the BH-host mass relations, there will be some 
BHs of these smaller masses living in larger halos which have already crossed 
the ``quenching'' mass, but the age distribution will still be one-sided and 
weighted to very young ages. To match the observed age trend, there 
must, in short, be some process which 
can trigger quasar activity at other halo masses {\em before} 
they cross the ``quenching'' threshold.

\begin{figure}
    \centering
    \figzoom
    \plotone{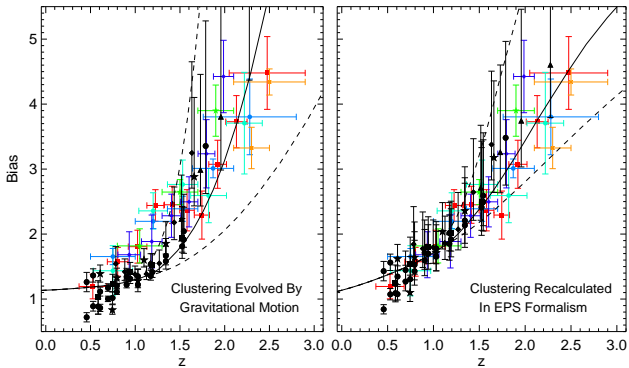}
    \caption{Observed clustering of quasars, compared to that inferred from 
    their $z=0$ early-type hosts if the termination of star formation is coincident 
    with quasar activity (as lower left panel in Figure~\ref{fig:age.bias.compare}). 
    {\em Left:} Our standard methodology is used to empirically evolve the clustering 
    of local systems (black points) to the redshifts shown. 
    {\em Right:} Instead, using the full EPS formalism and estimated $b(M_{\rm halo},\,z)$ 
    to evolve the clustering of local systems. Differences owing to the choice of 
    methodology are small at the halo masses of interest.
    \label{fig:compare.methods}}
\end{figure}

Finally, we note that 
in evolving the clustering of local systems ``up'' in redshift in the lower 
panels of Figure~\ref{fig:age.bias.compare}, there might be some ambiguity (if, for example, 
a given $z=0$ halo is assembled from many progenitor high-redshift 
halos with significantly different properties). The simple evolution predicted by 
Equation~(\ref{eqn:bias.evol}) is derived from pure gravitational motions, and therefore 
as applied moving ``backwards'' in time represents an effective ``mean'' bias 
of the progenitors of the $z=0$ system \citep[see][]{Fry96}. To the extent, however, that 
there is a dominant progenitor halo at a given redshift and many smaller
halos which will be accreted by the ``main'' halo, it is the properties  
of the main progenitor which are of interest here. 

We therefore consider a completely independent approach to empirically 
compare the clustering measurements shown, which attempts to capture 
these subtleties. Given a $z=0$ population, we can estimate its characteristic 
host halo mass either directly from the measurements of \citet{Mandelbaum06},  
or indirectly by matching the observed bias (with bias as a function of 
halo mass calculated for the adopted cosmology following \citet{MoWhite96} 
and \citet{ShethTormen01} as in \S~\ref{sec:populations}). Following 
\citet{Neistein06}, we then calculate the mass of the main progenitor
halos of this $z=0$ mass, 
as a function of redshift (i.e.\ the highest-mass ``branch'' of the EPS merger 
tree at each redshift). At the redshift of interest (e.g.\ appropriate lookback time, 
for the comparisons in the lower panels of Figure~\ref{fig:age.bias.compare}), 
we then calculate the expected bias for halos of this main progenitor mass. 

Figure~\ref{fig:compare.methods} reproduces the lower-left 
comparison in Figure~\ref{fig:age.bias.compare} (the expected clustering of elliptical 
progenitors at the times determined by their stellar population ages),  
using both our previously adopted methodology and this revised estimation. 
The latter method has the advantage, as noted above, of accounting for the difference 
between the main progenitor and smaller, accreted systems. The approach, however, suffers from 
certain inherent ambiguities in Press-Schechter theory. For example, the 
calculated evolution is not necessarily time-reversible, and the clustering properties 
are assumed to be a function of halo mass alone, which recent high-resolution 
numerical simulations suggest may not be correct 
\citep[e.g.,][]{Gao06,Harker06.bias,Wechsler06}. 
In particular, if quasars are triggered in mergers 
(i.e.\ have particularly recent halo assembly times for their post-merger halo masses), 
then they may represent especially biased regions of the density distribution. 
Unfortunately, it is not clear how to treat this in detail, as there remains considerable 
disagreement in the literature as to whether or not a significant ``merger bias'' exists 
\citep[see, e.g.][]{KH02,Percival03,Furlanetto06}. 
Furthermore the distinction between galaxy-galaxy and 
halo-halo mergers (with the considerably longer timescale for most galaxy mergers) 
means that it is not even clear whether or not, after the galaxy merger, there would be a 
significant age bias. In any case, most studies suggest the effect is quite small: using 
the fitting formulae from \citet{Wechsler02,Wechsler06}, we find that even in extreme cases 
(e.g.\ a $M\gg M_{\rm vir}$ halo merging at $z=0$ as opposed to an ``average'' 
assembly redshift $z_{f}\approx6$) the result is that the ``standard'' EPS formalism 
underestimates the bias by $\approx30\%$. For the estimated quasar host halo masses 
and redshifts of interest here, the maximal effect is $\lesssim 10\%$ at all $z=0-3$, 
much smaller than other systematic effects we have considered. 
This is consistent with \citet{Gao06} and \citet{Croton07} who find that ``assembly bias'' is only important 
(beyond the $10\%$ level) for the most extreme halos or galaxies in their simulations. 

In practice, Figure~\ref{fig:compare.methods} 
demonstrates that, for the halo masses of interest here, the two methods yield very similar 
results. This is reassuring, and owes to the fact that the differences 
from the choice of methodology discussed above are 
important only at very high or very low halo masses, where for example the clustering 
of small halos which are destined to be 
accreted as substructure in clusters ($\gtrsim 10^{15}\,h^{-1}\,M_{\sun}$) will be 
very different from the clustering of similar-mass halos in field or void environments. 
Alternatively, one can think of the EPS approach as attempting to account for 
the possibility that bias is a non-monotonic function of mass \citep[e.g.\ 
rising galaxy bias at very low luminosities,][]{Norberg02}, 
which Figure~\ref{fig:local.cluster.compare} demonstrates is important only at 
masses well below those of interest here. 
To the extent that any ``merger bias'' is permanent or long-lived 
(as expected if the excess clustering is correlated with halo concentration or 
formation time), our ``standard'' methodology should account for it, as we 
simply evolve the clustering of the present hosts of quasar ``relics'' to earlier times 
according to gravitational motions. That the different seen is small provides 
a further reassurance that the effects of ``merger bias'' are probably not dramatic. 
Ultimately, we have re-calculated all the results herein 
adopting the more sophisticated (but more model-dependent and potentially 
more uncertain) EPS approach, and find that it marginally improves the significance of our 
conclusions but leaves them qualitatively unchanged.

\section{Implications for High-Redshift Clustering}
\label{sec:redshift}

At $z\gtrsim2-3$, comparing quasar and early-type clustering becomes more ambiguous. 
Above $z\sim2$, the QLF ``turns over'', and the density of bright quasars declines. 
Specifically, it appears that the characteristic quasar luminosity $L_{\ast}$ 
declines \citep[][and references therein]{HRH06}, at least from 
$z\sim2$ to $z\sim4.5$ above which the ``break'' $L_{\ast}$ can no longer be determined. 
One possible interpretation of this is an extension of our analysis 
for $z\lesssim2$; i.e.\ one could assume that each ``quasar'' episode here signals 
the end of a BH's growth, which will evolve passively to $z=0$. At $z=0$, the 
tightness of the local BH-host relations means the hosts must have the appropriate mass 
and lie within the appropriate halos, to within a factor $\sim2$ of the observed scatter. 
Therefore, we can adopt the same approach as in \S~\ref{sec:populations} 
to use the local observed clustering as a function of host properties to 
evolve back in time and predict quasar clustering as a function of redshift. 

\begin{figure*}
    \centering
    \figzoom
    \plotone{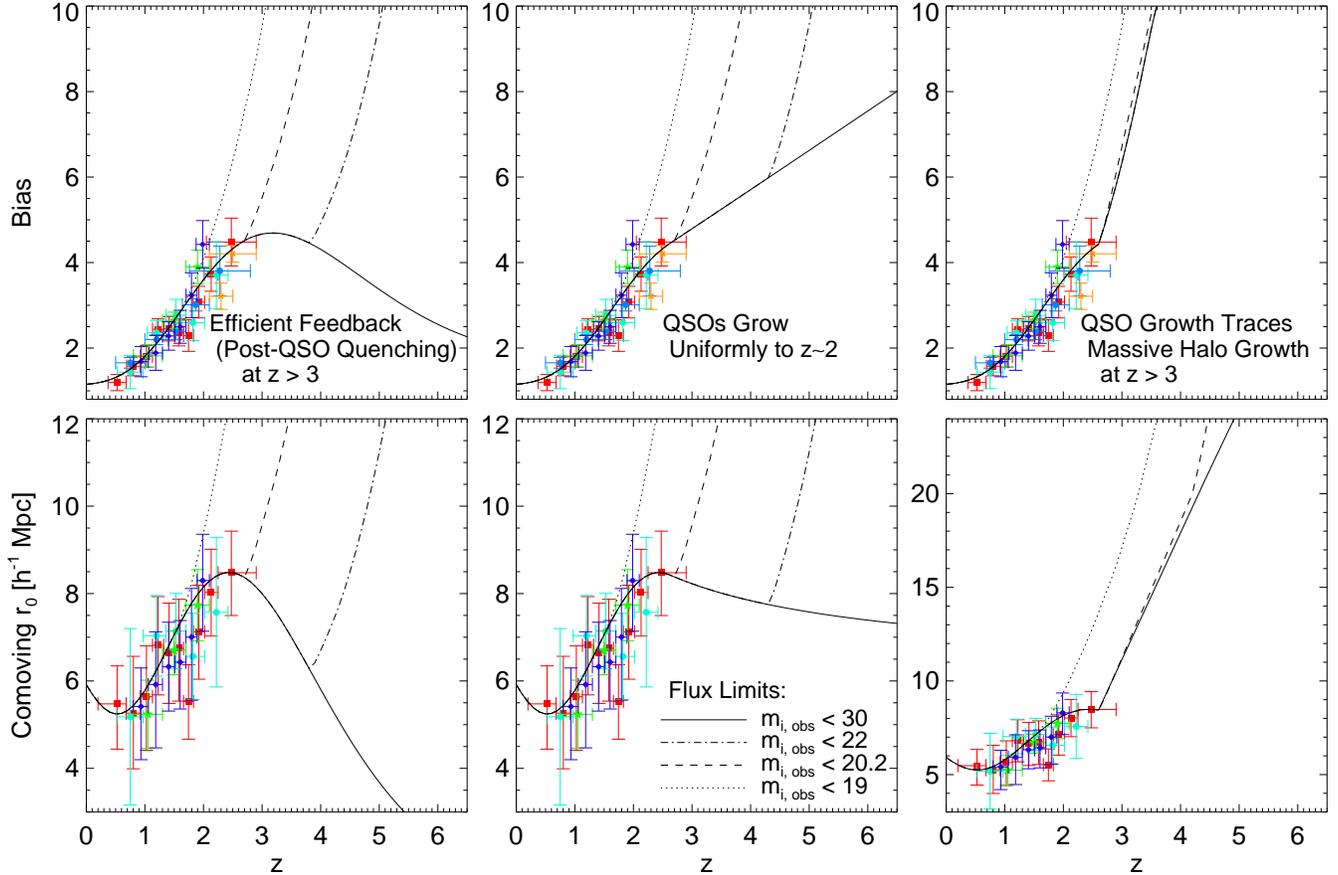}
    \caption{Using the model in Figures~\ref{fig:local.cluster.compare} \& \ref{fig:bias.populations} 
    to predict the observable clustering of quasars at high redshifts $z\gtrsim3$. Observed clustering is 
    shown (points) as in Figure~\ref{fig:local.cluster.compare}. Different lines show the 
    effect of different observed-frame $i$-band flux limits, as labeled 
    \citep[note $i=20.2$ corresponds to the SDSS DR3 completeness limit;][]{Richards06}. 
    Left panels assume efficient feedback at high redshifts; i.e.\ that BH growth 
    ``shuts down'' after each quasar episode. Center panels assume all $z>2$ BHs 
    grow with the observed QLF to the characteristic peak luminosities 
    at $z\sim2$, then shut down (``inefficient feedback''). 
    Right panels assume quasar growth tracks host 
    halo growth, even after a quasar episode, until $z=2$ (``maximal growth''). Future observations 
    at $z\sim4$ with moderately improved flux limits $m_{i}<22$ should be able 
    to break the degeneracies between these models. 
    \label{fig:highz}}
\end{figure*}

\begin{figure}
    \centering
    \figzoom
    \plotone{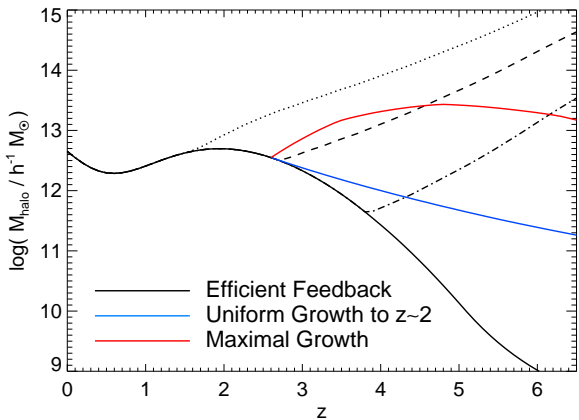}
    \caption{Characteristic inferred quasar-hosting 
    halo mass corresponding to the model clustering as a function of redshift 
    shown in Figure~\ref{fig:highz}, for our adopted cosmology. Dotted, dashed, and 
    dot-dashed lines are for the appropriate flux limits as in Figure~\ref{fig:highz}. 
    The black, blue, and red lines show the left, center, and right models 
    (``efficient feedback'', growth to typical $z\sim2$ quasars, and ``maximal'' growth, 
    respectively) from 
    Figure~\ref{fig:highz} with effectively infinitely deep flux limits ($i<30$); 
    all are identical below $z\sim2.6$. 
    \label{fig:mhalo}}
\end{figure}

Figure~\ref{fig:highz} shows the bias and correlation length predicted by 
this approach, an extension of the model (Equation~\ref{eqn:bestfit.bias}) 
we have considered at $z\lesssim2$. Figure~\ref{fig:mhalo} also shows the typical 
host halo mass corresponding to the predicted clustering as a function of redshift (for our 
adopted cosmology); in this simple extension of the $z\lesssim3$ case, the observed 
decline in the QLF $L_{\ast}$ traces a decline in the characteristic (although not most massive) 
quasar-hosting halo mass. 

However, at high redshifts, flux limits may severely bias clustering measurements. 
Although at $L<L_{\ast}$, quasar clustering does not strongly depend on 
the quasar luminosity (see \S~\ref{sec:bias.vs.lum}), 
implying a well-defined characteristic active mass which we can adopt \citep[see also][]{Lidz06}, 
this is not necessarily true for $L>L_{\ast}$. 
In fact, Figure~\ref{fig:bias.vs.lum} shows 
\citep[and observations may begin to see, e.g.][]{Porciani06} 
a steepening of bias versus luminosity at 
$L>L_{\ast}$, reflecting the uniformly high observed 
quasar Eddington ratios 
\citep{Vestergaard04,MD04,Kollmeier05} at high luminosities, which 
imply the bright end of the QLF ($L\gg L_{\ast}$) becomes predominantly 
a sequence in active BH mass. To the extent that BH mass traces 
host mass, then, these systems reside in more massive hosts and will be more 
strongly biased. 

In order to estimate how this will change 
the observed clustering, we roughly approximate this effect as follows. 
For a given flux limit at a given redshift there is a reasonably well-defined 
survey depth, to a minimum luminosity $L_{\rm min}$. If this is sufficiently 
deep to resolve the QLF ``break,'' i.e.\ $L_{\rm min}<L_{\ast}$, then 
the weak observed dependence of clustering on luminosity means the observed 
clustering will trace that characteristic of $\sim L_{\ast}$ quasars -- 
corresponding to characteristic $\mbh$ active BHs and $\mgal\approx\mbh/\mu$ hosts 
(our fiducial model, and the case for all observations plotted in Figure~\ref{fig:highz}). 
However, if the flux limit or redshift is sufficiently high 
such that $L_{\rm min}>L_{\ast}$, then the survey will not 
sample these characteristic host masses. In this case, we consider the 
bias as a function of luminosity plotted in Figure~\ref{fig:bias.vs.lum} 
from the models of \citet{H06b,Lidz06}, evaluated at $L_{\rm min}$ at the 
given redshift. Qualitatively, for the nearly constant 
Eddington ratios observed at $L>L_{\ast}$, $L\propto M_{\rm BH} \propto \mgal$, 
we expect $L_{\rm min}>L_{\ast}$ to correspond to an approximate 
minimum observed host mass, 
$M_{\rm min} \sim \mgal(L_{\ast}) \times (L_{\rm min}/L_{\ast}) \sim M_{\rm Edd}(L_{\rm min})/\mu$. 
Since the QLF slope is steep at $L>L_{\ast}$, objects near 
$L_{\rm min}$ or $M_{\rm min}$ will dominate the observed sample, 
and so this amounts to calculating the clustering for this mass, instead of $\mgal(L_{\ast})$, 
at the given redshift. We caution that this is a rough approximation to 
more realistic selection effects, but should give 
us some idea how flux limits will bias the observed clustering. 

We consider several representative flux limits, in observed-frame $i$-band, 
typical of optical quasar surveys (e.g.\ the SDSS), in addition to the 
case with effectively infinite depth ($m_{i}<30$). We calculate $L_{\ast}$ at the 
appropriate rest-frame wavelengths as a function of redshift using the 
fits to $L_{\ast}$ from \citet{HRH06}, spanning $z\sim0-6$ and spanning the relevant rest-frame 
wavelength intervals. 
At the limits of most current 
optical surveys, $m_{i}<20.2$, the QLF break $L_{\ast}$ is only 
marginally resolved at $z\sim2-3$, and so above this redshift 
surveys are systematically biased to more massive $L>L_{\ast}$ BHs and 
higher clustering amplitudes. However, a relatively modest improvement 
in depth to $m_{i}<22$ would allow unbiased clustering estimates 
to be extended to $z\sim4$. 

We have so far assumed BHs effectively ``shut down'' after their quasar epoch 
-- i.e.\ ``efficient feedback'' even at high redshifts.  
Although the various observations discussed above (Eddington ratio distributions, 
quasar host measurements, HOD models, black hole mass functions, and 
our clustering comparison) demand this be true at $z\lesssim2-3$, 
there are no such constraints at $z\gtrsim3$. 
In other words, it is possible that the increase in the QLF from $z\sim6$ to $z\sim3$ 
traces the growth of the same populations of BHs, not the subsequent triggering and 
``shutdown'' of different populations. 
If BHs at $z\sim6$ continue to grow to $z\sim2-3$ before 
``shutting down,'' they must live in more massive $z=0$ host galaxies 
(to preserve the tight observed BH-host mass relation), and thus should 
have stronger clustering amplitudes. 
We therefore consider two representative simple models which 
bracket the range of possibilities for this growth and present simple 
tests for future clustering measurements to break the 
degeneracy between these models. 

First, we assume that quasars grow with the QLF to $z\sim2$ before 
``shutting down'' (i.e.\ ``inefficient feedback''). In such a case, $z\sim6$ quasars grow either 
continuously or episodically with their host systems until the epoch where 
``downsizing'' begins, and the QLF at all redshifts $z>2$ represents the 
{\em same} systems building up hierarchically. The $z=0$ relic masses (and therefore 
$z=0$ characteristic host masses, from which we calculate the 
``parent'' halo clustering as a function of redshift) are then 
the same at all $z\ge2$.
This is also equivalent to a ``pure density evolution'' model for 
the high redshift QLF, as in \citet{Fan01}, in which 
the QLF break luminosity $L_{\ast}$ remains constant 
above $z\sim2-3$ while the number density/normalization 
uniformly declines. Such a model is marginally disfavored by 
current measurements \citep{HRH06}, but constraints on $L_{\ast}$ 
at high redshifts are sufficiently weak that it remains a possibility. 

The clustering as a function of redshift in this model behaves very differently 
from the previous model at high redshifts. If objects cannot grow after 
their quasar epoch even at high redshifts, then the subsequent decline of the QLF $L_{\ast}$ 
traces a decline in characteristic active masses, and the bias of 
active systems ``turns over''; however, if all grow to the characteristic 
$L_{\ast}$ at $z\sim2$, then these high redshift systems all represent similar $z=0$ masses 
by the time they ``shut down,'' and must be increasingly biased at higher redshifts. 

Next, we consider a ``maximal'' growth model, in which we assume not only 
that the buildup of the QLF represents the continued growth of BHs until $z\sim2$, 
but also that this growth proportionally tracks the typical growth of 
dark matter halos over this redshift range. We very crudely estimate the 
``typical growth'' with the growth of an average high-redshift quasar ``host'' halo. 
Based on their space density and BH mass, \citet{Fan01,Fan03} estimate 
that typical $z\sim5-6$ SDSS quasars represent $\sim6\sigma$ 
overdensities. We therefore assume that quasars at a given 
redshift $z>2$, with a typical $L_{\ast}$ and corresponding $M_{\rm BH}$ at 
that redshift, will grow by the same proportional amount as a halo which represents 
a $6\sigma$ fluctuation 
\citep[characteristic of halos hosting observed $z\sim6$ quasars,][]{YLi06} 
from the observed redshift to $z=2$, after which growth ``shuts down.'' This then yields 
the $z=0$ BH mass, corresponding host mass, and evolved clustering. 
Note that, although similar, this is not the same as assuming quasars track 
$6\sigma$ overdensities at $z>2$, because to the extent that the QLF $L_{\ast}$ 
does {\em not} grow by the same proportionality, this model effectively allows 
``new'' or different BHs/host halos to dominate the QLF at different redshifts. 
It simply mandates that they all grow at this rapid rate. 
For example, an observed $z\sim6$ BH of $\sim10^{8}\,M_{\sun}$ is assumed to 
reach a mass of $2\times10^{9}\,M_{\sun}$ 
at $z=2$ (and then shuts down, so that this is also the 
mass at $z=0$), and a $\sim10^{8}\,M_{\sun}$ observed quasar 
at $z=4$ will grow to $\sim5\times10^{8}\,M_{\sun}$. 
The choice of rate 
is arbitrary, we choose it as a reasonable upper limit. In any case, the 
predicted evolution of the bias as a function of redshift is extremely steep, 
so the exact values will be very sensitive to the growth model and 
adopted cosmology. The point we wish to illustrate is that 
this model generically predicts a steep bias evolution at 
$z>2$, which regardless of the details will be distinguishable 
if future quasar clustering measurements at $z=3-4$ can be extended to a 
depth of $m_{i}<22$. 

Note that extending the depth of quasar surveys to $m_{i}<22$ will 
move further down the QLF and increase the 
density of quasars observed, meaning a smaller survey can be used to 
constrain the clustering to comparable accuracy as the SDSS or 2dF. 
Using the \citet{HRH06} QLF to estimate the relevant space density of 
quasars above the flux limit as a function of redshift and assuming the 
errors in clustering amplitude relative to those in \citet{Croom05} 
scale as $N_{\rm qso}^{-1/2}$, we estimate 
that for the redshift interval $3.5<z<4.5$ ($3.75<z<4.25$) 
a field size of $\sim25\,{\rm deg}^{2}$ ($50\,{\rm deg}^{2}$) 
would be sufficient to distinguish between the first two models 
(efficient high-redshift feedback and all high-redshift quasars growing 
to $z\sim2$ luminosities) at $\sim2.5-3\,\sigma$. The last model (``maximal 
evolution'') predicts an even more extreme departure in clustering 
properties, and could be distinguished or ruled out at $\sim2.5-3\,\sigma$ 
by clustering observations from $2.75<z<3.25$ in just a 
$\sim8-15\,{\rm deg}^{2}$ field.

\section{Discussion/Conclusions}
\label{sec:discussion}

We compare the clustering of quasars and different galaxy populations
as a function of morphology, mass, luminosity, and redshift, and 
demonstrate that these comparisons can be used to robustly 
rule out several classes of models for quasar triggering and 
the association between quasar and galaxy growth. In each case, 
the observations favor a model which associates quasars 
with the ``formation event'' of ellipticals, a strong prediction of 
theoretical models which argue that major, gas-rich mergers 
form ellipticals and trigger quasar activity \citep{H06b}. 

The predicted bias as a function of mass/luminosity 
for systems which once hosted quasars agrees well at all masses 
and luminosities with that observed for early-type populations. 
In other words, the clustering of an $\mgal$ elliptical galaxy is 
exactly what we would expect if these galaxies, which typically contain 
$M_{\rm BH}=\mu\,\mgal$ \citep[$\mu\sim0.001$][]{Magorrian98,MarconiHunt03,HaringRix04} 
BHs, represented the dominant ``hosts'' of the quasar population for a brief 
period, setting $L_{\ast}$ at that 
redshift with an Eddington-limited $L_{\ast}=L_{\rm Edd}(M_{\rm BH})$ ``epoch'' 
of activity. 
In the most basic sense, this is a confirmation that ellipticals today were 
indeed the host population of high-redshift quasars \citep[see also][]{Porciani04,Croom05}, 
with the appropriate 
corresponding BH masses. This should not be surprising, since 
the \citet{Soltan82} argument demonstrates that most BH mass must have been 
accumulated in bright, near-Eddington ``quasar'' epochs, 
and the tightness of the local 
BH-host mass relation \citep[and similar BH mass-host property relations, see][]{Novak06} 
argues that BH growth must be tightly coupled to the host properties. 
However, there are additional non-trivial implications.

First, this implies that there really is a characteristic host and BH 
mass ``active'' at a given epoch, traced by the QLF $L_{\ast}$. This is an
important prediction of certain theoretical models for quasar lightcurves 
\citep{H05c, H05d}, 
and supported by other lines of observation above. Furthermore, 
this implies that the formation ``epoch'' for BHs of a given mass 
must be relatively short in time, as continually adding BHs of a given 
mass (at lower Eddington ratio or in radiatively inefficient states) to 
the population would dilute the agreement in Figure~\ref{fig:local.cluster.compare}. 
Quasars are active in characteristically different parent halo populations at 
different redshifts -- i.e.\ most systems cannot undergo multiple separate 
periods of quasar activity, at least at $z\lesssim2$.
We find further support for this by considering 
observed quasar clustering as a function of luminosity, 
which favors the predictions of \citet{Lidz06}, namely a relatively weak 
trend of bias as a function of luminosity. In fact, the combination of quasar clustering 
measurements as a function of luminosity and redshift supports at high 
significance previous suggestions of little or no luminosity dependence 
\citep[e.g.,][]{Adelberger05.lifetimes,Myers06,Myers06b}, and is inconsistent with 
the predictions of simplified ``light bulb'' or exponential quasar light curve models 
\citep[e.g.,][]{KH02,WyitheLoeb03} at $>4-5\,\sigma$.

This relates to a subtle but important distinction: this implies 
that the halos of the dominant population of 
$\mgal$ ellipticals are {\em the same halos} as those which 
hosted the corresponding quasar activity. A significant fraction of the $\mgal$ 
early-type population cannot form from later collapsing 
halos, as this both requires the buildup of BHs of the same mass at a 
different time, ruled out by the observations above, and would dilute the 
clustering agreement. 

Second, the clustering of late-type galaxies at a given 
luminosity or mass does {\em not} agree with the evolved 
clustering of quasars. This argues that it is specifically the progenitors of 
{\em early-type} galaxies which hosted quasars. Although this is 
not surprising, given that observations find it is specifically {\em bulge} 
mass/velocity dispersion which correlates with BH mass \citep[][]{Gebhardt00,FM00,MD02,HaringRix04}, 
there still exist classes of models which would 
{\em generically} associate BH formation with galaxy formation, star formation, or halo virialization. 
Our comparison of observations rules out these scenarios.

We further invert our comparisons to predict quasar clustering as a 
function of redshift, and compare this with the observed clustering of 
red and blue galaxies at each redshift. Quasars do not trace a uniform/constant 
population with redshift -- i.e.\ they are not cosmologically long-lived, 
as has been noted in many previous clustering studies \citep[see][]{Porciani04,Martini04,Croom05}. 
Further, quasars do not trace the clustering of ``established'' red or blue galaxy 
populations. This rules out models in which quasars are associated with 
cyclic \citep[e.g.,][]{CiottiOstriker01,Binney04} or radio 
``heating'' modes over a Hubble time in red galaxies, as well 
as (at least the most straightforward implementations of) 
models which generically associate quasars with star formation 
\citep[e.g.,][]{Granato04} or disk instabilities in high-redshift, 
gas-rich disk systems \citep[e.g.,][]{KH00}. 

Note, however, that this does not rule out the presence of these accretion 
modes at low luminosities and/or low redshifts. 
Many of the observations discussed above limit bright, high Eddington 
ratio quasar activity to a single, short-lived epoch. Long-lived 
accretion in a ``radio-mode'' is believed to be associated 
with particularly low Eddington ratio activity \citep[][]{Ho02,White06}, 
perhaps an entirely different accretion state 
\citep[][]{NY95,Marchesini04,Jester05,Pellegrini05,Koerding06}, and 
not typical $L_{\ast}$ QLF activity. Models which invoke ``radio mode'' type 
accretion at low Eddington ratios in quasar ``relics'' \citep[e.g.,][]{Croton05,deLucia06} 
are therefore completely consistent with the clustering arguments herein. 
Other observations of 
black hole mass functions and quasar Eddington ratios \citep[][]{YT02,MD04,Kollmeier05}
also rule out ``cyclic'' models, insofar as they attempt to explain 
more luminous $\sim L_{\ast}$ quasar activity. 
However, theoretical models of stochastic, feedback-regulated accretion in 
gas-rich systems \citep{HH06} predict that these fueling modes 
dominate at low redshift and at typical ``Seyfert'' luminosities at higher 
redshifts, even where mergers \citep{H06c,H06d} may dominate the 
bright $\sim L_{\ast}$ quasar population. We demonstrate that the 
luminosities at which these ``Seyfert'' accretion modes may dominate 
are sufficiently low that they have no effect on our results. However, 
we make predictions for future measurements of clustering as a function of luminosity at 
moderate redshifts, which may be able to detect such changes in the 
characteristic host population and fueling mechanisms through 
the cross-correlation of galaxies and faint, X-ray selected AGN in 
deep fields at $z\sim1$.

It is also important to distinguish the processes which may be associated 
with the initial formation of ellipticals from their subsequent 
evolution. Once morphologically transformed by a gas-rich merger, for example, 
mass can be moved ``up'' the red sequence (galaxies increased in mass)
by gas-poor mergers, which will involve neither star formation nor 
quasar activity, but it cannot be {\em added} to the red 
sequence in this manner. As noted above, low-luminosity AGN 
or ``radio mode'' activity, or halo shock ``quenching'' 
may be of critical importance to suppressing cooling flows and further accretion in 
massive ellipticals. However, these do not appear to be associated with the initial 
formation of an elliptical or triggering of traditional, bright, high-redshift quasars. 

It is interesting that, at all redshifts, quasars clustering is observed to 
be intermediate between blue and red galaxy clustering. 
To the extent that 
halos grow monotonically with time, intermediate clustering 
may imply intermediate halo mass and therefore, perhaps, an 
intermediate evolutionary stage -- i.e.\ quasars are representative 
of an {\em evolutionary} state ``between'' blue and red galaxies. 
This is predicted in many theoretical models \citep[][]{Granato04,
Scannapieco04,SDH05a,Monaco05,H06b,H06d}, which 
posit that quasar feedback, or a merger-triggered quasar phase, is 
associated with (regardless of whether or not it directly {\em causes}) 
the formation of ellipticals and ``transition'' of galaxies to the red sequence. 
Quasars do appear to cluster similarly to likely ``merger'' populations 
(close pairs, post-starburst/E+A galaxies, and sub-millimeter galaxies)
at both low ($z\lesssim0.3$) and high ($z\sim2-3$) redshifts, but 
more detailed measurements of these populations are needed.
We confirm the finding from previous studies 
\citep{Porciani04,Croom05,Porciani06,Myers06,Myers06b} that quasar 
clustering is consistent with a constant halo mass $\sim4\times10^{12}\,h^{-1}\,M_{\sun}$; 
interestingly, 
similar to the characteristic mass of small galaxy groups (at least at low redshift) 
in which galaxy-galaxy major mergers are expected to be most efficient. Further 
theoretical investigation of this, and the possibility that it may manifest in 
an excess of quasar clustering on small scales 
\citep[e.g.,][but see also Myers et al.\ 2007b]{Hennawi06}, are important subjects for future work. 
We calculate the implied time delay between the star-forming/LBG ``phase'' 
of evolution, quasar phase, and red galaxy phase, from the evolved clustering 
of halos of a given initial mass. Although this is just a ``toy'' model of 
self-similar halo evolution, suggestively, the time from quasar epoch to 
``red galaxy'' epoch corresponds reasonably at all redshifts with the delay 
expected for stellar populations to redden to typical 
red galaxy colors after the termination or ``shutdown'' of star formation. We caution, however, 
that the systematic uncertainties in measurements of these clustering strengths 
(especially at high redshifts) remain a concern, and future studies which compare 
uniformly selected galaxy and quasar populations across a wide range of redshifts 
adopting consistent measurement methods are needed to make these conclusions 
robust. 

We explore this further by considering the age of BHs (i.e.\ the time 
since the mean ``formation'' 
or quasar epoch for BHs of a given relic mass) as a function of mass, and comparing 
this with the age of their hosts. The mean stellar age 
(and dispersion in ages) of early-type BH hosts 
agrees well at all masses, implying that quasars are associated with the 
formation epoch of early-type galaxies. Specifically, these are light-weighted or 
single-burst ages of red galaxies, which tend to reflect the last significant epoch of 
star formation -- i.e.\ quasars are associated with the last significant epoch (or 
potentially the termination) of star formation in elliptical hosts. 
This association can be used to accurately predict quasar clustering as a function of 
redshift in a purely empirical manner, without any assumptions about 
quasar light curves, lifetimes, or Eddington ratios. 

A similar association does not hold for disk-dominated systems hosting BHs; i.e.\ 
again demonstrating that 
BH growth and quasar activity do not generically trace star formation. 
Likewise, such an association does not hold for dark matter halos, 
meaning that quasars 
do not generically trace halo formation/assembly 
\citep[even accounting for the halo downsizing effects seen in][]{Neistein06}. 
They also do not generically trace the crossing of host halos of the 
critical shock ``quenching'' mass in \citet{Birnboim03,Keres05,Dekel06}.
Although our comparisons are consistent with the possibility that this halo mass 
does ``quench'' gas accretion (and thus quasar activity will not generally 
occur at higher masses), 
there must be a mechanism which can trigger quasars before they cross 
this threshold.

We emphasize that this does {\em not} imply that most of the stars in spheroids 
form in such a short-lived burst contemporaneous with their 
quasar epoch. Direct calculation of the inferred stellar population 
ages from line index and SED fitting \citep[following][]{Trager00} 
for realistic star formation histories 
from the semi-analytic models of \citet{Somerville01} and hydrodynamical 
merger simulations of \citet{Robertson06b} suggests that the 
ages inferred for present early-type galaxies indeed reflect the epoch of 
the termination of star formation, even when $\gtrsim95\%$ of stars are 
formed over a much longer timescale at significantly earlier times 
(in these cases, in quiescent star formation in disks). Indeed, 
most models \citep[e.g.,][]{H06e,Croton05,deLucia06} predict that only a small 
fraction of stars in ellipticals were formed in merger-induced starbursts. 
This, combined with the 
lack of a general correlation between star formation in disks and quasar 
activity, supports the hypothesis that quasars trace the {\em end} of star 
formation in present spheroids, as predicted by models which associate 
quasar activity with mergers (or other mechanisms) that rapidly exhaust 
gas and transform (assemble) disks into spheroids. 

Finally, we extend these empirical predictions of quasar clustering to 
high redshift, and show that the $z>2-3$ clustering of quasars is 
dramatically different depending on whether or not 
feedback is efficient at high redshift (i.e.\ whether or not 
$z>2$ quasars ``shut down'' after their quasar episode, as observations 
show they do at $z<2$). Present 
observations cannot distinguish these possibilities, but 
future quasar samples at $z\sim3-4$ with flux limits $m_{i}\lesssim22$ 
should be able to break the degeneracies. 

Although there are non-negligible uncertainties in these conversions 
to characteristic host masses or luminosities, a number of different lines of evidence 
support their robustness. 
The conversion to a characteristic BH mass (and the lack of 
evolution to $z=0$) is determined by direct Eddington ratio 
observations \citep{Heckman04,Vestergaard04,MD04,Kollmeier05,HHN06} 
and observed cosmic background constraints \citep{ERZ02,Ueda03,Cao05}. 
If we instead adopt the simplest empirically 
inferred models of quasar lightcurves from matching the QLF and 
local BH mass function \citep{YT02,Marconi04,Shankar04,HRH06},
we come to an identical conclusion.
Likewise for more sophisticated models which incorporate 
effects of different accretion states and compare radio, X-ray, and 
optical QLFs \citep{Merloni04.bhmf}, and theoretical models of quasar light curves 
from numerical hydrodynamic simulations of galaxy mergers 
which dynamically incorporate BH growth and feedback 
\citep{H06b}. Whether we adopt the observed BH-host mass/luminosity 
relations, or those from simulations \citep{DiMatteo05,Robertson06a,H05a,H05b,H05c}, 
we obtain an identical result. Adopting just the BH-host mass relation, 
and using the time since the quasar epoch to determine the observed $M/L$ (i.e.\ 
assuming this represents a stellar age and using the population synthesis 
models of \citet{BC03} to predict $M/L$ in a given band) also does not change this comparison.

Finally, direct construction of halo occupation (HOD) models 
from observed quasar clustering 
\citep{Porciani04,Porciani06,Adelberger05.hostmass,Croom05} leads to the same conclusions
regarding the luminosity $L_{\ast}$ reflecting the evolving characteristic mass 
of active quasars, their weak subsequent BH growth, and host properties at $z=0$. 
In other words, our uncertainties in this approach are most likely dominated 
by the substantial measurement errors in the bias $b(z)$ of quasar and galaxy populations, 
not by the systematics in our methodology. As such, future improved measurements 
of quasar clustering and bias at high redshifts, particularly 
as a function of luminosity (e.g. using the proximity effect; Faucher-Giguere
et al. 2006), as 
well as improved galaxy clustering measurements which can resolve 
the clustering as a function of mass or luminosity at $z\gtrsim1$, will 
strengthen the constraints herein and continue to inform
models for quasar fueling and their associations with spheroid formation.

\acknowledgments We thank 
Michael Strauss for helpful discussions, and are grateful to the anonymous 
referee whose comments improved this paper. 
This work was supported in part by NSF grant AST
03-07690, and NASA ATP grants NAG5-12140, NAG5-13292, and NAG5-13381.

\end{document}